\documentclass[aps,reprint,amsmath,superscriptaddress,prb,showpacs,amsfonts,twocolumn]{revtex4-1}

\pdfoutput=1

\usepackage{epsfig}
\usepackage{amsmath}
\usepackage{amssymb}
\usepackage{bm}
\usepackage{color}
\usepackage{graphicx}

\newcommand{\beq}{\begin{equation}}
\newcommand{\eeq}{\end{equation}}
\newcommand{\beqarray}{\begin{eqnarray}}
\newcommand{\eeqarray}{\end{eqnarray}}
\newcommand{\Ham}[1][]{\ensuremath{{\cal{H}}_{\text{#1}}}} 
\newcommand{\eq}[1]{Eq.~(\ref{#1})} 
\newcommand{\fig}[1]{Fig.~(\ref{#1})} 
\newcommand{\Sec}[1]{Sec.~\ref{#1}} 
\newcommand{\Ref}[1]{Ref.~\onlinecite{#1}} 
\newcommand{\tab}[1]{Table~\ref{#1}} 

\begin{document}

\allowdisplaybreaks

\title{Charge and spin supercurrents in triplet
      superconductor--ferromagnet--singlet superconductor Josephson
      junctions} 
\author{P. M. R. Brydon}
\email{brydon@theory.phy.tu-dresden.de}
\affiliation{Institut f\"{u}r Theoretische Physik, Technische Universit\"{a}t
  Dresden, D-01062 Dresden, Germany}
\author{Wei Chen}
\affiliation{Max-Planck-Institut f\"{u}r Festk\"{o}rperforschung,
  Heisenbergstrasse 1, D-70569 Stuttgart, Germany}
\author{Yasuhiro Asano}
\affiliation{Department of Applied Physics, Hokkaido University, Sapporo,
  060-8628, Japan}
\author{Dirk Manske}
\affiliation{Max-Planck-Institut f\"{u}r Festk\"{o}rperforschung,
  Heisenbergstrasse 1, D-70569 Stuttgart, Germany}
\affiliation{Yukawa Institute for Theoretical Physics, Kyoto University, Kyoto
  606-8502, Japan} 

\begin{abstract}
We study the Josephson effect in a triplet
superconductor--ferromagnet--singlet superconductor junction. We show that the
interaction of tunneling Cooper pairs with the interface magnetization can
permit a Josephson current at the lowest order of a tunneling Hamiltonian
perturbation theory. Two conditions must be satisfied for this to occur: the
magnetization of the ferromagnet has a component parallel to the ${\bf
  d}$-vector of the triplet superconductor, and the gaps of the
superconductors have the same parity with respect to the interface
momentum. The resulting charge current displays an unconventional dependence 
on the orientation of the magnetic moment and the phase difference. This is
accompanied by a phase-dependent spin current in the triplet superconductor,
while a phase-independent spin current is always present.
The tunneling perturbation theory predictions are confirmed using a 
numerical Green's function method. An analytical treatment of
a one-dimensional junction demonstrates that our conclusions are robust far
away from the tunneling regime, and reveals signatures of the unconventional
Josephson effect in the critical currents. 
\end{abstract}

\date{May 2, 2013}


\maketitle

\section{Introduction}

The physics of spin-triplet superconductors (TSs) is much richer than their
spin-singlet superconductor (SS) counterparts due to the spin degree of
freedom of a triplet Cooper pair. Although a bewildering variety of triplet
pairing states are in principle possible for any crystal
symmetry,~\cite{SigUed1991} there are only a handful of systems where a TS
state has been well established, e.g. UPt$_3$ and
Sr$_2$RuO$_4$.~\cite{JoyTai2002,Norman2011,MacMae2003,Maeno2012} Even in these  
cases, many questions remain about the precise form of the TS order 
parameter.~\cite{KalBer2009,Norman2011,Maeno2012} Much effort has therefore
been directed at developing experimental tests capable of unambiguously
identifying TS. A promising route is to incorporate the candidate TS into a
heterostructure device and search for signatures of the odd-parity orbital
state in tunneling measurements.~\cite{tunspec,TSDN} Alternatively, the spin
part of the Cooper pair wavefunction can be probed by bringing the TS into
contact with a ferromagnet (FM). This is manifested by the crucial role of the 
relative orientation between the vector order parameters of the TS and the FM,
the ${\bf d}$-vector and magnetization respectively, in controlling the
physics of the
device.~\cite{tunspecTSFM,Annunziata2011,TSFM,TSFMprox,FTFspinvalve}

Theoretical investigations of the Josephson effect between TSs have revealed
many remarkable consequences of their intrinsic spin structure. For example,
the spin of the Cooper pair permits the existence of a Josephson spin current
between TSs with misaligned ${\bf d}$-vectors,~\cite{asanospin,LinSudFMTS}
similar to the spin supercurrent observed in superfluid
$^3$He.~\cite{fomin,bunkov} Another notable proposal is the triplet
superconductor-ferromagnet-triplet superconductor (TFT)
junction,~\cite{Kastening06,Brydon08,BryMan2009,BujTimBry2012} where the 
coupling of the Cooper pairs' spin with the exchange field in the 
barrier causes a sign reversal of the current as the orientation of the
exchange field with respect to the ${\bf d}$-vectors is varied. This is in
stark contrast to the well-known $0$-$\pi$ transition in magnetic junctions
between SSs,~\cite{GolKupIli2002,Eschrig2003,SSFMreviews} which is independent
of the orientation of the barrier magnetization. 
Such an anomalous $0$-$\pi$ transition would therefore be extremely strong
evidence of the triplet state of the superconductors. On the other hand,
creating a TFT junction from any of the TS candidate materials is
experimentally challenging  due to the high purity requirements needed for the
TS. Even with recent success in growing superconducting thin films of
Sr$_2$RuO$_4$,~\cite{krockenberger} such devices will likely remain
hypothetical for some time, due to the greater challenges posed by growing a
layered heterostructure. More immediately plausible is to create a triplet
superconductor-ferromagnet-singlet superconductor (TFS) junction by coating
the Sr$_2$RuO$_4$ thin film with a ferromagnetic layer and then contacting to
a conventional superconductor.

For junctions between superconductors of like parity, the lowest harmonic in
the Josephson current vs phase difference $\phi$ relation is usually
$\sin(\phi)$, which originates from tunneling processes involving only a 
single Cooper pair. This term is necessarily absent in a nonmagnetic 
Josephson junction between an SS and a TS, however, due to the 
orthogonal spin pairing states.~\cite{PalHaeMaa1977} Instead, the
singlet-triplet 
conversion can only occur in processes involving the coherent tunneling of
even numbers of Cooper pairs, and so $\sin(2\phi)$ is the leading harmonic in
the current vs phase relation.~\cite{Yip1993,KwoSenYak2004,LuYip2009}
The coupling between a single tunneling Cooper pair and the magnetic degrees of
freedom in a magnetically-active barrier, on the other hand, can accomplish
the conversion between singlet and triplet spin states,~\cite{Eschrig2003}
hence generating a lowest-order Josephson coupling in the sense discussed
above. An example of such a magnetic interaction is the intrinsic spin-orbit
coupling expected to occur at the junction
interface.~\cite{Fenton,GeshLark1986,MilRaiSau1988,AsaTanSigKas,ZutMaz2005}    
This has been proposed as the origin of the unexpectedly large Josephson
currents in junctions between single crystals of Sr$_2$RuO$_4$ and
conventional $s$-wave 
SS, and the pronounced dependence on the crystal face upon which the Josephson
contact is made.~\cite{Jin2000,Nelson2004,Liu2010,Saitoh2012}  
In contrast, relatively little work has been done for a ferromagnetic
tunneling barrier. Previous studies have found a highly unconventional
Josephson charge current in a TFS junction which is \emph{even} in the phase
difference and \emph{odd} in the 
component of the magnetization parallel to the ${\bf d}$-vector of the
TS.~\cite{TanKas1999,TanKas2000,YokTanGol2007} The origin of this current,
and the conditions under which it occurs, nevertheless remain obscure.

\begin{figure}
\includegraphics[width=\columnwidth]{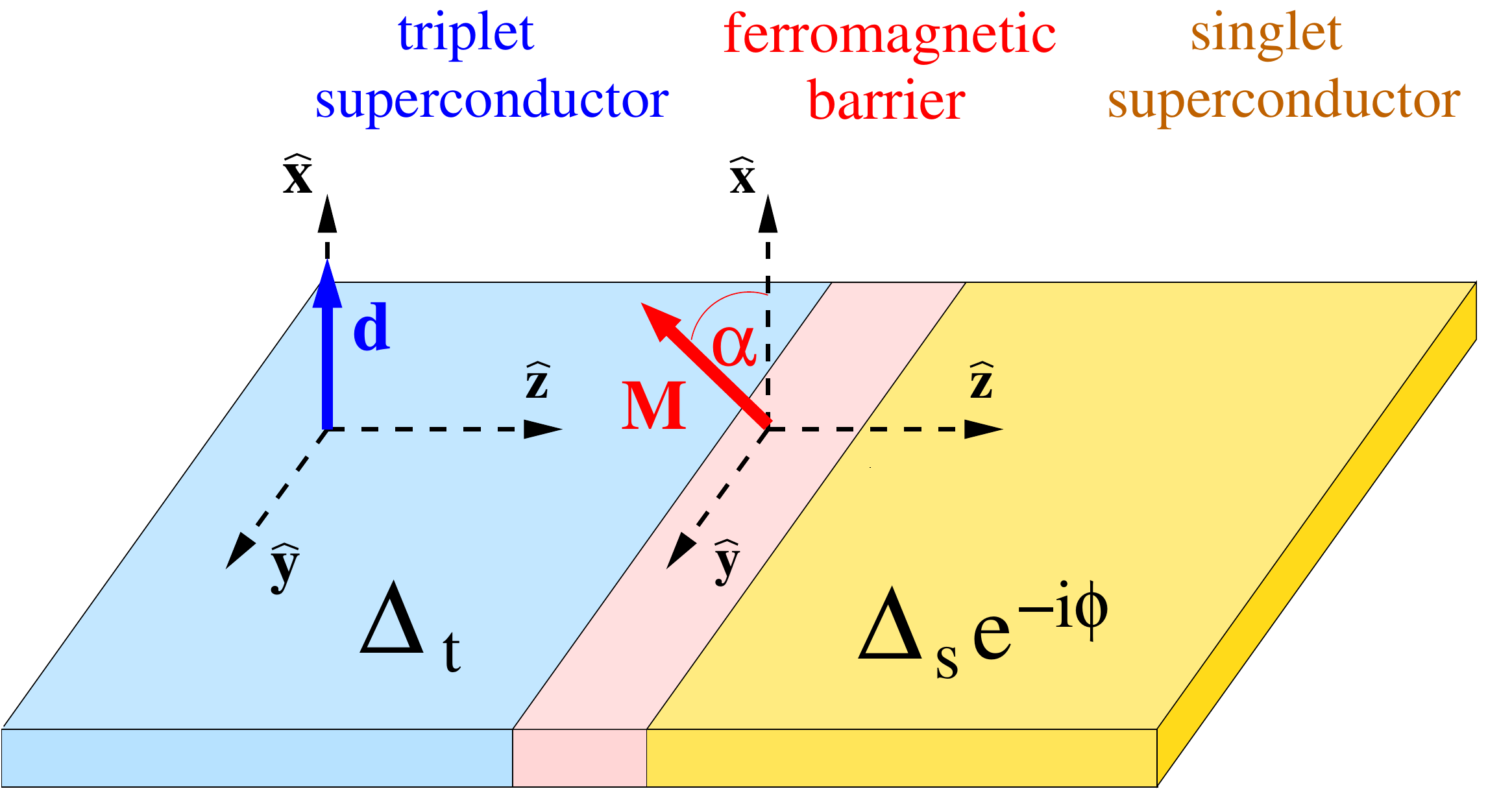}
\caption{\label{fig:schematic}{(color online) Schematic diagram of the
  Josephson junction considered in this paper. The ${\bf d}$-vector of
    the TS defines the $x$-axis, and we restrict the magnetization to the
    $x$-$y$ plane, making the angle $\alpha$ to the $x$ axis. The phase
    difference between the TS and SS is given by 
    $\phi$. }}
\end{figure}

It is the purpose of this paper to perform a detailed theoretical study of the
Josephson effect in a TFS 
junction, shown schematically in~\fig{fig:schematic}. Similar to the TFT
junction, we find highly unconventional Josephson physics which originates
from the coupling of the barrier moment to the spin of the
tunneling quasiparticles comprising the Cooper pairs. Using a tunneling
Hamiltonian perturbation  theory, we show that there is a lowest-order
Josephson  charge current when the orbital pairing states of the superconductors
have the same parity with respect to the interface momentum, and the
magnetization has a component parallel to the ${\bf 
  d}$-vector of the TS. There is
also a spin current in the TS, which has both a phase-dependent and a
phase-independent contribution. The latter is a universal spin supercurrent
which appears at triplet superconductor-ferromagnet interfaces and is due to
spin-dependent reflection processes.~\cite{TSFM} We test the predictions of
the perturbation theory  
using both lattice and continuum models of the junction. In the lattice
theory we survey a wide selection of different orbital pairing states in the
SS and TS. The focus of the continuum theory, on the other hand, is to
understand the role of resonant tunneling through the Andreev bound states at
the junction interface. Both approaches yield excellent
agreement with the perturbative analysis at sufficiently high
temperatures. Although 
deviations from perturbation theory become more severe as the temperature is
lowered, it nevertheless remains qualitatively correct down to zero
temperature. 

This paper is organized as follows. In~\Sec{sec:perturbation} we present a
perturbative theory for the Josephson effect in the TFS. The predictions of
this section are  confirmed first in~\Sec{sec:lattice} by numerically
determining the currents in a microscopic lattice model of the junction, and
then by analytical calculation for a continuum model
in~\Sec{sec:quasiclassics}. Our concluding discussion is given
in~\Sec{sec:conclusions}.

\section{Perturbation Theory} \label{sec:perturbation}

\subsection{Hamiltonian of the junction}

The Hamiltonian of the TFS junction is written
\beq
\Ham = \Ham[TS]+ \Ham[SS] + \Ham[tun] + \Ham[ref]\,. \label{eq:pt:Ham}
\eeq
Here $\Ham[TS]$ and $\Ham[SS]$ describe the bulk TS and SS on either
side of the barrier. We have 
\beqarray
\Ham[TS] & = &  \frac{1}{2}\sum_{\bf k}\Psi^{\dagger}_{t,{\bf
    k}}\left(\begin{array}{cc} 
\epsilon_{t,\bf{k}}\hat{\sigma}_{0} &
i{\bf d}_{\bf k}\cdot{\hat{\pmb{\sigma}}}\hat{\sigma}_{y} \\
(i{\bf d}_{\bf k}\cdot{\hat{\pmb{\sigma}}}\hat{\sigma}_{y})^{\dagger} &
-\epsilon_{t,{\bf{k}}}\hat{\sigma}_{0}
\end{array}
\right)\Psi^{}_{t,\bf{k}}\,, \label{eq:Hamt} \\
\Ham[SS] & = & \frac{1}{2}\sum_{\bf k}\Psi^{\dagger}_{s,{\bf
    k}}\left(\begin{array}{cc} 
\epsilon_{s,\bf{k}}\hat{\sigma}_0 &
i\Delta_{s,\bf k}e^{-i\phi}\hat{\sigma}_{y} \\
-i\Delta^{\ast}_{s,\bf k}e^{i\phi}\hat{\sigma}_{y} &
-\epsilon_{s,{\bf{k}}}\hat{\sigma}_0
\end{array}
\right)\Psi^{}_{s,\bf{k}}\,, \notag \\\label{eq:Hams}
\eeqarray
where 
$\Psi^{}_{\nu,\bf k} = (c^{}_{\nu,{\bf k},\uparrow}, c^{}_{\nu,{\bf
    k},\downarrow}, c^{\dagger}_{\nu,-{\bf k},\uparrow},
c^{\dagger}_{\nu,-{\bf k},\downarrow})^{T} 
$ and $c^{}_{\nu,{\bf k},\sigma}$ ($c^{\dagger}_{\nu,{\bf k},\sigma}$) are the
fermion annihilation (creation) operators for a spin-$\sigma$ quasiparticle
with momentum ${\bf k}$ in the $\nu=s$ ($t$) SS (TS). The bare
dispersion in  each superconductor is $\epsilon_{\nu,{\bf k}}$. We only
consider unitary equal-spin-pairing states for the TS. Without loss of 
generality, it is convenient to let the vector order 
parameter of the TS define the $x$-axis, i.e. ${\bf d}_{\bf
  k}=\Delta_{t,{\bf k}}{\bf e}_{x}$. Other orientations of ${\bf d}_{\bf k}$
can be achieved by spin rotation of the system, and do not result in new
physics. The global phase difference between the TS and the SS is
$\phi$.

The TS and SS are connected by the tunneling Hamiltonian, which we define
\beq
\Ham[tun] = \sum_{\nu=s,t}\sum_{{\bf k},{\bf
    k}'}\sum_{\sigma,\sigma'}T^{\sigma,\sigma'}_{\nu,{\bf k},{\bf
    k}'}c^{\dagger}_{\overline{\nu},{\bf k},\sigma}c^{}_{\nu,{\bf k}',\sigma'}\,, \label{eq:Hamtun}
\eeq
where $\overline{\nu}=s (t)$ for $\nu=t (s)$. In order to properly model the
interaction of the tunneling quasiparticles with the magnetic moment of the
barrier we require explicitly 
spin-dependent tunneling matrix elements. Following~\Ref{BryMan2009},
we also include a reflection Hamiltonian 
\beq
\Ham[ref] = \sum_{\nu = s,t}\sum_{{\bf k},{\bf
    k}'}\sum_{\sigma}R^{\sigma,-\sigma}_{\nu,{\bf k},{\bf
    k}'}c^{\dagger}_{\nu,{\bf k},\sigma}c^{}_{\nu,{\bf k}',-\sigma}\, , \label{eq;Hamref}
\eeq
to properly account for the interaction of the quasiparticles with the FM
layer. We only include spin-flip reflection processes in $\Ham[ref]$, since 
spin-preserving reflection processes clearly do not contribute to
either the charge or spin currents. 

\subsubsection{Ansatz for tunneling and reflection matrix elements}

Our perturbation analysis crucially relies upon the form of the tunneling and
reflection matrix elements. In particular, it is necessary to include the
phase shift acquired by the quasiparticles during the various tunneling or
reflection processes.~\cite{BryMan2009}
Although it is in principle possible to determine the tunneling and reflection
matrix elements from a more fundamental Hamiltonian, here we motivate a
phenomenological form by comparison with an exactly-solvable scattering
problem. 

A common model for the tunneling barrier is a $\delta$-function
potential,~\cite{TanKas2000,BryMan2009,SenYak2008,LuYip2009,KwoSenYak2004,Kastening06,Brydon08} i.e $\hat{V}({\bf r}) =
U_0\hat{\sigma}_0\delta(z) + {\bf U}_{M}\cdot\hat{\pmb \sigma}\delta(z)$, where $U_0$
is the charge scattering potential and ${\bf U}_M =
U_M\left(\cos\alpha\hat{\bf e}_x+\sin\alpha\hat{\bf e}_y\right)$ is the
magnetic scattering potential. It is straightforward to evaluate the
scattering matrix for this potential, 
and hence determine the transmission and reflection coefficients,
$t_{\nu,\sigma,\sigma'}({\bf k},{\bf k}')$ and 
$r_{\nu,\sigma,\sigma'}({\bf k},{\bf k}')$ respectively. The
  transmission and spin-flip reflection coefficients are both small when the
  magnetic potential is large, i.e. the dimensionless parameter
  $g=k_{F}|{\bf U}_M|/2E_F$ is much larger than one, where $k_F$ is the Fermi
  wavevector and $E_F$ is the Fermi energy. It
is reasonable  to expect that in this limit we should 
have $T^{\sigma,\sigma'}_{\nu,{\bf k},{\bf k}'}\sim 
t_{\nu,\sigma,\sigma'}({\bf k},{\bf k}')$ and $R^{\sigma,-\sigma}_{\nu,{\bf k},{\bf k}'}\sim
r_{\nu,\sigma,-\sigma}({\bf k},{\bf k}')$. 
By requiring that a tunneling or
reflected quasiparticle acquires the same phase as in the exact solution, we
hence have the following ansatz for the matrix elements
\begin{subequations} \label{eq:pt:matel1}
\beqarray
T^{\sigma,\sigma}_{\nu,{\bf k}, {\bf k}'} &=& \frac{1}{g^2}T^{sp}(k_z,k_z')\delta_{{\bf k}_\parallel,{\bf k}_\parallel'}\,,\\
T^{-\sigma,\sigma}_{\nu,{\bf k}, {\bf k}'} &=&
\frac{i{\nu}e^{i\sigma\alpha}}{g}T^{sf}(k_z,k_z')\delta_{{\bf k}_\parallel,{\bf k}_\parallel'}\,,\\
R^{-\sigma,\sigma}_{\nu,{\bf k}, {\bf k}'} &= & \frac{i{\nu}e^{i\sigma\alpha}}{g}R^{sf}(k_z,k_z')\delta_{{\bf k}_\parallel,{\bf k}_\parallel'}\,,
\eeqarray
\end{subequations}
where $\nu=-1$ $(+1)$ as a factor for $\nu=t$ $(s)$ as a subscript.
  Crucially for our analysis, the phase shift acquired during spin-flip
  tunneling or reflection depends on the initial spin $\sigma$ and the angle
  $\alpha$ of the magnetic moment in the $x$-$y$ plane
[see~\fig{fig:schematic}].
We assume 
that the $T^{sp}(k_z,k_z')$, $T^{sf}(k_z,k_z')$, and
$R^{sf}(k_z,k_z')$ are real functions, independent of $g$, and satisfy
\begin{subequations} \label{eq:pt:matel2}
\begin{gather}
 T^{sp}(k_z,k_z') = T^{sp}(-k_z,-k_z') = T^{sp}(k_z',k_z), \\
T^{sf}(k_z,k_z') = -T^{sf}(-k_z,-k_z') = T^{sf}(k_z',k_z) \,, \\
R^{sf}(k_z,k_z') = -R^{sf}(-k_z,-k_z') = -R^{sf}(k_z',k_z) \,.
\end{gather}
\end{subequations}
These conditions originate from both the comparison to the
scattering coefficients and the requirement that $\Ham[tun]$ and $\Ham[ref]$
be Hermitian. Note that the $\delta_{{\bf k}_\parallel,{\bf k}_\parallel'}$
in~\eq{eq:pt:matel1} ensures the conservation of momentum parallel to the
barrier, a consequence of the translational invariance along the
interface.~\cite{BruOttZim1995}  

Although we have motivated our ansatz~\eq{eq:pt:matel1} by comparison with the 
$\delta$-function barrier, we expect our approach to be of more general
validity.~\cite{BujTimBry2012} In particular, the spin-dependent phase shifts
acquired by the tunneling quasiparticles should be robust to other choices of
barrier model, since they are a consequence of the orientation of the
magnetization.

\subsection{Perturbation theory}

The number operator for particles in each spin sector of the two
superconductors is given by
$N_{\nu,\sigma}=\sum_{\bf{k}}c^{\dagger}_{\nu,{\bf{k}},\sigma}c^{}_{\nu,{\bf{k}},\sigma}$. From 
this we define the associated particle currents
\beq
I^\nu_{\sigma}=-\nu\langle\partial_{t}N_{\nu,\sigma}\rangle \,.
\eeq
We proceed by expanding the $S$ matrix to lowest order in
${\cal H}^\prime = \Ham[tun]+\Ham[ref]$, hence treating the
tunneling and reflection processes as a perturbation of the Hamiltonian
${\cal H}_{0} = \Ham[TS] + \Ham[SS]$.~\cite{Mahan,BryMan2009} This is
justified so long as the tunneling and reflection matrix elements are
small, which by our ansatz~\eq{eq:pt:matel1} holds in the $g\gg1$
limit. By the Kubo formula, we then have 
\beq
I^\nu_{\sigma}=-i\nu\int^{t}_{-\infty}dt'\langle\left[\partial_{t}N_{\nu,\sigma}(t),
  \Ham^\prime(t') \right]\rangle\,, \label{eq:pt:curnusig}
\eeq
where the time-dependence is given within the interaction picture, i.e.
${\cal O}(t) = e^{i{\cal H}_{0}t}{\cal O}e^{-i{\cal H}_{0}t}$, with 
  ${\cal H}_0 = K_{0} + 
\sum_{\nu}\mu_{\nu}\sum_{\sigma}N_{\nu,\sigma}$ and $K_{0}$ is the
associated grand canonical Hamiltonian. Since we are only 
interested in the DC Josephson effect, we take the same chemical potential
in both the TS and SS, i.e. $\mu_{s}=\mu_{t}=\mu$. We hence find for the
terms in the commutator in~\eq{eq:pt:curnusig}
\beqarray
\partial_{t}N_{\nu,\sigma}(t) &=& i\left\{B^{-\sigma,\sigma}_{\nu}(t) -
  B^{\sigma,-\sigma}_{\nu}(t)\right\} \notag \\
&& + i\sum_{\varsigma}\left\{A^{\varsigma,\sigma}_{\nu}(t) -
  A^{\sigma,\varsigma}_{\overline{\nu}}(t)\right\}\,, \\
{\cal H}^\prime(t) & = & \sum_{\nu}\sum_{\varsigma}B^{\varsigma,-\varsigma}_{\nu}(t) + 
\sum_{\nu}\sum_{\varsigma,\varsigma'}A^{\varsigma,\varsigma'}_{\nu}(t)\,,
\eeqarray
where we introduce the operators
\beqarray
A^{\varsigma,\varsigma'}_{\nu}(t) &=
&\sum_{\bf{k},\bf{k}'}T^{\varsigma,\varsigma'}_{\nu,\bf{k},\bf{k}'}c^{\dagger}_{\overline{\nu},{\bf{k}},\varsigma}(t)c^{}_{\nu,{\bf{k}}',\varsigma'}(t)\,,
\\
B^{\varsigma,\varsigma'}_{\nu}(t) &=
&\sum_{\bf{k},\bf{k}'}R^{\varsigma,\varsigma'}_{\nu,{\bf{k}},{\bf{k'}}}c^{\dagger}_{\nu,{\bf{k}},\varsigma}(t)c^{}_{\nu,{\bf{k}}',\varsigma'}(t) \,.
\eeqarray
The time-dependence of the fermion operators in these expressions
is given by $c^{}_{\nu,{\bf k},\sigma}(t) = 
e^{iK_{0}t}c^{}_{\nu,{\bf k},\sigma}e^{-iK_{0}t}$.

Following standard arguments,~\cite{BryMan2009,Mahan,LinSudFMTS} we write
the current~\eq{eq:pt:curnusig} as
\beq
I^\nu_{\sigma} = 2\nu\text{Im}\left\{\Phi^{\text{ret}}_{\nu,\sigma}(\omega=0) +
\Psi^{\text{ret}}_{\nu,\sigma}(\omega=0)\right\}\,,
\eeq
where the retarded correlation functions
$\Phi^{\text{ret}}_{\nu,\sigma}(\omega)$ and
$\Psi^{\text{ret}}_{\nu,\sigma}(\omega)$ give the contributions from tunneling
and reflection processes, respectively. They are obtained by analytic
continuation $i\omega_n\rightarrow\omega + i0^{+}$ of the corresponding
Matsubara functions
\beqarray
\Phi_{\nu,\sigma}(i\omega_n) & = & \int^{\beta}_{0}d\tau
e^{i\omega_n\tau}\sum_{\varsigma,\varsigma',\varsigma''}\langle T_{\tau}
A^{\varsigma,\sigma}_{\nu}(\tau)A^{\varsigma',\varsigma''}_{\nu}(0)\rangle \,,
\\
\Psi_{\nu,\sigma}(i\omega_n) & = &\int^{\beta}_{0} d\tau
e^{i\omega_n\tau}\langle T_{\tau} B^{-\sigma,\sigma}_{\nu}(\tau)
B^{-\sigma,\sigma}_{\nu}(0) \rangle \,.
\eeqarray
The Matsubara functions are evaluated by using Wick's theorem to expand the
two-particle correlators. We hence
find the particle currents in the TS
\beqarray
I^t_{\sigma} & = & -\frac{1}{g^2}\sum_{{\bf k},{\bf
    k}'} \left[R^{sf}(k_z,k_z')\right]^2\delta_{{\bf k}_\parallel,{ \bf
    k}_\parallel' }\notag \\
&& \times \mbox{Im}\left\{e^{2i\sigma\alpha} \frac{\Delta^{\ast}_{t,{\bf
      k}}\Delta_{t,{\bf k}'}}{E_{t,{\bf k}}E_{t,{\bf
      k}'}}\right\}F_{t,t}({\bf k},{\bf k}') \notag \\
& & -\frac{2}{g^3} \sum_{{\bf k},{\bf
    k}'}T^{sp}(k_z,k_z')T^{sf}(k_z,k_z')\delta_{{\bf
    k}_\parallel,{\bf
    k}_\parallel'} \notag \\
&& \times \mbox{Re}\left\{e^{i(\phi+\sigma\alpha)}\frac{\Delta^\ast_{s,{\bf k}}\Delta_{t,{\bf
      k}'}}{E_{s,{\bf k}}E_{t,{\bf k}'}}\right\}F_{s,t}({\bf k},{\bf k}')\,, \label{eq:pt:tripcur}
\eeqarray
and the SS
\beqarray
I^s_{\sigma} & =& \frac{2\cos(\alpha)}{g^3}\sum_{{\bf k},{\bf
    k}'}T^{sp}(k_z,k_z')T^{sf}(k_z,k_z')\delta_{{\bf
    k}_\parallel,{\bf
    k}_\parallel'} \notag \\
&& \times \mbox{Re}\left\{e^{i\phi}\frac{\Delta^\ast_{s,{\bf k}}\Delta_{t,{\bf
      k}'}}{E_{s,{\bf k}}E_{t,{\bf k}'}}\right\}F_{s,t}({\bf k},{\bf
  k}')\,. \label{eq:pt:singcur}
\eeqarray
Here we utilize the convenient short-hand notation~\cite{LinSudFMTS}
\beqarray
F_{\nu,\nu'}({\bf k},{\bf k'})& = & \frac{f(E_{\nu,{\bf k}}) -
    f(E_{\nu',{\bf k}'})}{E_{\nu,{\bf k}} - E_{\nu',{\bf k}'}} \notag \\
&&  +
  \frac{1 - f(E_{s,{\bf k}}) - 
    f(E_{t,{\bf k}'})}{E_{s,{\bf k}} + E_{t,{\bf k}'}}\,,
\eeqarray
where
\beq
E_{\nu,{\bf k}} = \sqrt{(\epsilon_{\nu,{\bf k}}-\mu)^2 + |\Delta_{\nu,{\bf k}}|^2}\,,
\eeq
is the dispersion in the $\nu$ superconductor and $f(E)$ is the Fermi
distribution function. Equations~(\ref{eq:pt:tripcur}) and~(\ref{eq:pt:singcur})
are the central results of our perturbation theory analysis, as they give all
lowest-order contributions to the particle current in each
superconductor. Note that in the TS we have a contribution from reflection
processes, whereas in the SS only tunneling processes contribute to the
current. The lowest-order (i.e. single-Cooper-pair) tunneling processes
involve both a spin-preserving 
and a spin-flip tunneling event, which is necessary to transform the spin
singlet Cooper pairs of the SS into the $S_z=\pm\hbar$ triplet Cooper pairs
of the TS and \emph{vice versa}. The particle current in the
TS~[\eq{eq:pt:tripcur}]
depends on the spin $\sigma$ through the spin-dependent phase shifts acquired
during the scattering; in contrast, the particle current in the
SS~[\eq{eq:pt:singcur}] is 
the same for each spin orientation, as required by the singlet pairing state.

The important role of spin-flip tunneling implies a transfer of spin to the FM
tunneling barrier. In our calculation, however, we regard 
the magnetic moment of the FM barrier to have fixed magnitude and direction.
Including the response of the FM to the injected spin current is a challenging 
problem, requiring a nonequilibrium treatment that is
beyond the scope of the current manuscript. 

\subsection{Charge and spin currents}

\begin{table*}
\begin{ruledtabular}
\begin{tabular}{cccc}
TS gap symmetry & SS gap symmetry & charge current
[\eq{eq:pt:chargecurrent}] &   
spin current [\eq{eq:pt:spincurrent}]\\ \hline
$p_z$, $p_z+ip_y$ & $s$, $d_{y^2-z^2}$ & $I_{c}\cos(\alpha)\cos(\phi)$ & $I_{s,r}\sin(2\alpha) +
I_{s,t}\sin(\alpha)\sin(\phi)$ \\
$p_y$ & $s$, $d_{y^2-z^2}$ & 0 & $I_{s,r}\sin(2\alpha)$ \\
$p_z$ & $d_{yz}$ & 0 & $I_{s,r}\sin(2\alpha)$ \\
$p_y$& $d_{yz}$ & $I_{c}\cos(\alpha)\cos(\phi)$ &
$I_{s,r}\sin(2\alpha) + 
I_{s,t}\sin(\alpha)\sin(\phi)$ \\
$p_z+ip_y$ & $d_{yz}$ & $I_{c}\cos(\alpha)\sin(\phi)$ & $I_{s,r}\sin(2\alpha) +
I_{s,t}\sin(\alpha)\cos(\phi)$
\end{tabular}
\end{ruledtabular}
\caption{\label{tab:lowestorder}Table showing the Josephson charge and
  spin currents appearing in the lowest order of our perturbation theory
    for different combinations of TS and SS gap symmetries. In the interests
  of brevity we restrict ourselves to gaps lying in the $y$-$z$ plane. The
  Josephson current amplitudes are expected to satisfy
  $I_{c}\sim g^{-3}$, 
  $I_{s,r}\sim g^{-2}$, $I_{s,t}\sim g^{-3}$ in the
  tunneling limit $g\gg1$. The numerical values of these terms depend upon
  the details of the junction, e.g. the normal-state dispersion, the
    structure factor of the gaps, the properties of the tunneling region,
    etc.}   
\end{table*}

We can use either~\eq{eq:pt:tripcur} or~\eq{eq:pt:singcur} to calculate the
Josephson charge current $I_c =
-e(I^\nu_{\uparrow}+I^\nu_{\downarrow})$, as by charge conservation this is
the same in each superconductor. We hence find
\beqarray
I_c & = & 4e\frac{\cos(\alpha)}{g^3}\sum_{{\bf k},{\bf
    k}'}T^{sp}(k_z,k_z')T^{sf}(k_z,k_z')\delta_{{\bf k}_\parallel,{\bf
    k}_\parallel'} \notag \\
&& \times\text{Re}\left\{e^{i\phi}\frac{\Delta^{\ast}_{s,{\bf
      k}}\Delta_{t,{\bf 
      k}'}}{E_{s,{\bf k}}E_{t,{\bf k}'}}\right\}F_{s,t}({\bf k},{\bf k}')\,. \label{eq:pt:chargecurrent}
\eeqarray
The charge current strongly depends upon the orientation of the
magnetic moment through the $\cos(\alpha)$ factor. This implies that
  reversing the direction of the barrier moment also reverses the sign of the
  current, as was previously observed in~\Ref{TanKas2000}. The 
origin of this factor is the
interference of the particle currents in each spin sector of the
TS, which are phase-shifted with respect to one-another by $\pm2\alpha$
  as a consequence of the spin-flip tunneling, see the second term
  in~\eq{eq:pt:tripcur}. 

We extract the current vs phase relationship by examining the summand
in~\eq{eq:pt:chargecurrent}. In order to have a lowest-order
Josephson effect, we require that the product
$T^{sp}(k_z,k_z')T^{sf}(k_z,k_z')\Delta^{\ast}_{s,{\bf
    k}}\Delta_{t,{\bf k}'}\delta_{{\bf k}_\parallel,{\bf
    k}_\parallel'}$
not be odd in any component of ${\bf k}$ or ${\bf k}'$. 
From~\eq{eq:pt:matel2} we hence deduce
that the gaps $\Delta^{\ast}_{s,{\bf k}}$ and $\Delta_{t,{\bf
    k}'}$ have opposite parity with respect to the $z$-component of the
wavevector; equivalently, the gaps must have the  
same parity with respect to the interface momentum. For an $s$-wave SS,
therefore, there is a Josephson current
\beq
I_c \propto \frac{1}{g^3}\cos(\alpha)\cos(\phi)\,, \label{eq:pt:IJ}
\eeq
when the TS has $p_z$-wave symmetry. The proportionality
constant is determined by the details of the junction, such as
the normal-state dispersion or the structure of the interface. On the other
hand,~\eq{eq:pt:chargecurrent} is vanishing for a $p_y$-wave TS, and so a
Josephson current only appears in the next order of perturbation theory. The
current vs phase relation is then $I_c \propto 
\sin(2\phi)$, as for the nonmagnetic barrier.~\cite{PalHaeMaa1977} A
list  of the Josephson charge current vs phase relationships predicted
by~\eq{eq:pt:chargecurrent} for different 
combinations of orbital pairing states is given in~\tab{tab:lowestorder}.

The spin current is only present in the TS, as the singlet Cooper pairs
do not carry spin. The spin current is polarized along the
$z$-axis and is given by
\beqarray
I_{s,z} & = & \frac{\hbar}{2}\left(I^t_{\uparrow} -
I^t_{\downarrow}\right) \notag \\
& = & -\hbar\frac{\sin(2\alpha)}{g^2}\sum_{{\bf k},{\bf
    k}'} \left[R^{sf}(k_z,k_z')\right]^2\delta_{{\bf k}_\parallel,
    {\bf k}_\parallel'} \notag \\
&& \times \mbox{Re}\left\{\frac{\Delta^{\ast}_{t,{\bf
      k}}\Delta_{t,{\bf k}'}}{E_{t,{\bf k}}E_{t,{\bf
      k}'}}\right\}F_{t,t}({\bf k},{\bf k}') \notag \\
& & -\hbar\frac{2\sin(\alpha)}{g^3} \sum_{{\bf k},{\bf
    k}'}T^{sp}(k_z,k_z')T^{sf}(k_z,k_z')\delta_{{\bf
    k}_\parallel,{\bf
    k}_\parallel'} \notag \\
&& \times \mbox{Im}\left\{e^{i\phi}\frac{\Delta^\ast_{s,{\bf k}}\Delta_{t,{\bf
      k}'}}{E_{s,{\bf k}}E_{t,{\bf k}'}}\right\}F_{s,t}({\bf k},{\bf k}')\,. \label{eq:pt:spincurrent}
\eeqarray
The first term 
in~\eq{eq:pt:spincurrent} is the spin current due to spin-flip reflection from
the interface, where a spin $S_{z}=\pm\hbar$ Cooper pair is reflected as a
$S_{z}=\mp\hbar$ Cooper pair.~\cite{TSFM,BryMan2009,BujTimBry2012} As this
involves two 
spin-flip reflection events, the Cooper pair acquires a phase
shift of $\pm2\alpha$ and the magnitude of this term goes as $\sim
g^{-2}$. Furthermore, due to its origin in reflection processes, it is
sensitive to the orbital structure of the triplet gap, e.g. the sign reversal
of the $p_z$-wave gap upon specular reflection gives the Cooper pairs an
additional $\pi$
phase shift relative to the $p_y$-wave case, and there is
hence a sign difference in the respective reflection spin
currents. In contrast, the second term
in~\eq{eq:pt:spincurrent} originates from the  
interference of the spin-$\uparrow$ and -$\downarrow$ tunneling particle
currents, similar to the charge current, and goes as $\sim g^{-3}$. We
therefore expect 
that the reflection spin current dominates the tunneling spin current in
the tunneling limit $g\gg1$.
Whereas the reflection spin current is always present, the condition for the
tunneling 
term in~\eq{eq:pt:spincurrent} to be 
non-zero is the same as for a finite lowest-order charge current. Again
considering the case of 
an $s$-wave SS, we thus find that for a $p_z$-wave TS there is the spin
current 
\beq
I_{s,z} \propto \frac{\gamma}{g^2}\sin(2\alpha) +
\frac{1}{g^3}\sin(\alpha)\sin(\phi)\,, \label{eq:pt:ISzJ}
\eeq
where $\gamma$ is a numerical constant; for a $p_y$-wave TS, in contrast,
only the first term is present, and has opposite
sign. The Josephson spin currents are given
in~\tab{tab:lowestorder} for various combinations of orbital pairing states.

\begin{figure}[t!]
\includegraphics[width=\columnwidth,clip=true]{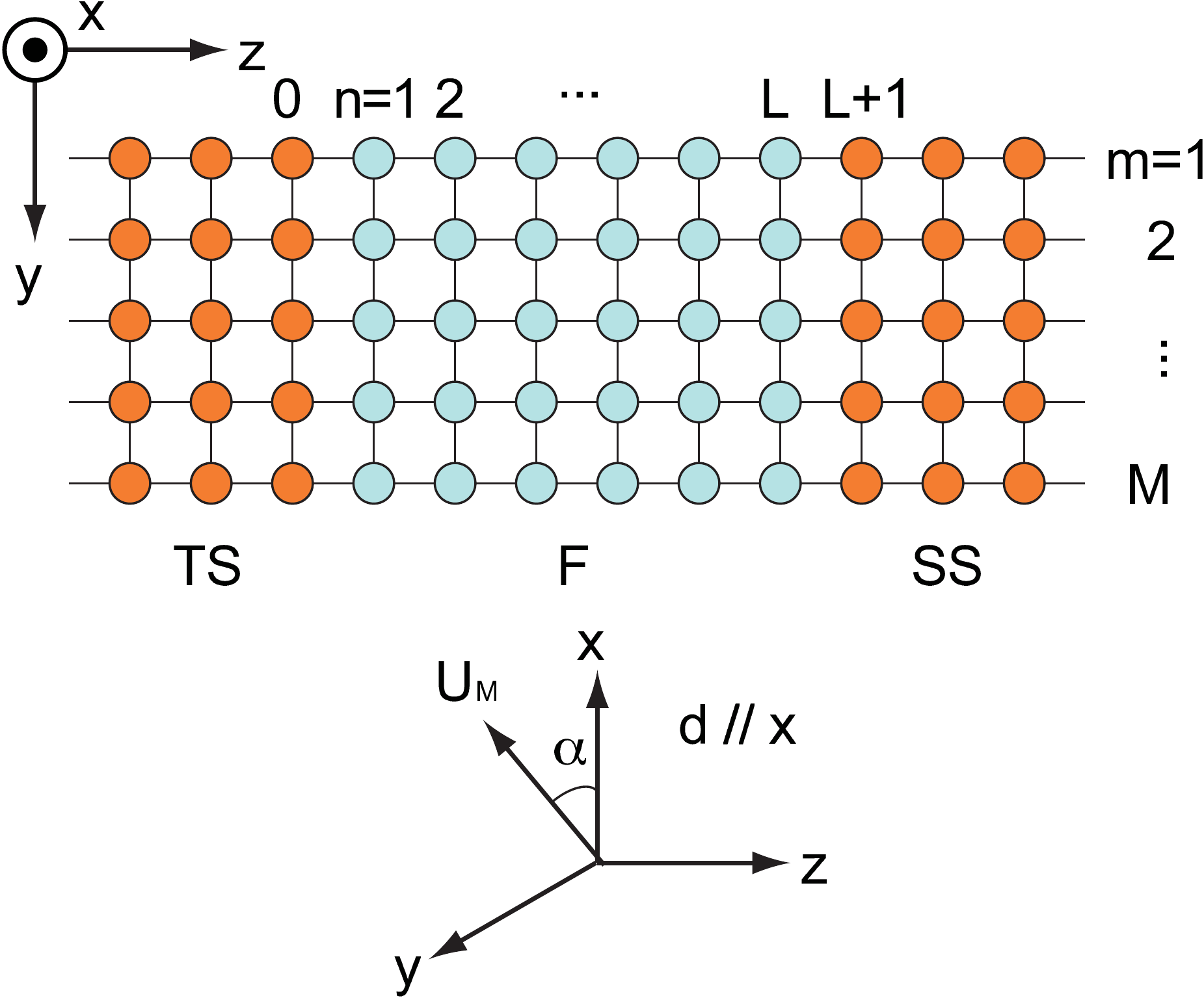}
\caption{(color online) Schematic diagram of the two-dimensional lattice model of the TFS
  junction. The length of the ferromagnetic layer and the junction width are
  $L$ and $M$ unit cells, respectively. The on-site magnetic potential ${\bf
    U}_M({\bf r})$ in the ferromagnetic layer is oriented in the $x$-$y$
  plane. }   
\label{asano1}
\end{figure}

\section{Lattice model of the junction} \label{sec:lattice}

In this section we test the predictions of the perturbation theory by
calculating the charge and spin Josephson currents in the tunneling regime of
a two-dimensional microscopic lattice model of the TFS junction.  Using the
recursive Green's function method,~\cite{Lee,asano01} we examine representative
examples of the different orbital combinations listed
in~\tab{tab:lowestorder}.

\subsection{Formulation}

We write the Hamiltonian describing the TFS junction in real space as
\beq
H= \frac{1}{2}
\sum_{{\bf r}, {\bf r}'}
\Psi^{\dagger}({\bf r})
\left(\begin{array}{cc}
\hat{H}_0({\bf r},{\bf r}')  & \hat{\Delta}({\bf r},{\bf r}')\\
-\hat{\Delta}^\ast({\bf r},{\bf r}') & - 
\hat{H}_0^\ast({\bf r},{\bf r}')
\end{array}\right)
\Psi^{}({\bf r}')\,, \label{bcs}
\eeq
where $\Psi({\bf r}) = (\psi^{}_\uparrow({\bf r}), \psi^{}_\downarrow({\bf r}),
\psi^\dagger_\uparrow({\bf r}), \psi^{\dagger}_\downarrow({\bf r}))^T$
is the vector of field operators and the sum in~\eq{bcs} is over the lattice
sites. For the normal state Hamiltonian $\hat{H}_0({\bf r},{\bf r}')$ we
assume a two-dimensional square lattice tight-binding model in the $y$-$z$
plane
\beqarray
\hat{H}_0({\bf r},{\bf r}') &= & \left[
- t\left\{ \delta_{{\bf r},{\bf r}'+{\bf z}}
+ \delta_{{\bf r},{\bf r}'-{\bf z}}
+\delta_{{\bf r},{\bf r}'+{\bf y}}
+\delta_{{\bf r},{\bf r}'- {\bf y}}\right\} \right. \notag \\
&& \left.+ \mu\delta_{{\bf r},{\bf r}'}
\right]\hat{\sigma}_0 +{\bf U}_M({\bf r}) \cdot \hat{\boldsymbol{\sigma}}
\delta_{{\bf r},{\bf r}'},
\eeqarray
The vectors are represented as ${\bf r}=n{\bf z}+m{\bf y}$ 
where ${\bf z}$ and ${\bf y}$ are the unit vectors of the 
tight-binding lattice in the $z$ and $y$ directions, respectively. 
In the $y$-direction, we apply the periodic boundary condition.
The exchange potential ${\bf U}_M({\bf r})$ is only non-zero in the ferromagnetic
barrier region of the junction. A sketch of the lattice model is shown
in~\fig{asano1}.
The pair potentials for the different pairing symmetries considered here have
the following real-space forms:
\begin{widetext}
\beq
\hat{\Delta}({\bf r},{\bf r}')=
\begin{cases} 
{\displaystyle \Delta e^{-i\phi} \delta_{\boldsymbol{r},\boldsymbol{r}'}
i \hat{\sigma}_y} & s\text{-wave},\\
{\displaystyle ({\Delta e^{-i\phi}}/{2}) 
i \hat{\sigma}_y
\left[-\delta_{{\bf r},{\bf r}'+{\bf z}+{\bf y}} 
- \delta_{{\bf r},{\bf r}'-{\bf z}-{\bf y}} 
+\delta_{{\bf r},{\bf r}'+{\bf z}-{\bf y}} 
+ \delta_{{\bf r},{\bf r}'-{\bf z}+{\bf y}} \right]}
 & d_{zy}\text{-wave},\\
{\displaystyle i({\Delta }/{2}) 
\hat{\sigma}_z
\left[\delta_{{\bf r},{\bf r}'+{\bf z}} - 
\delta_{{\bf r},{\bf r}'-{\bf z}} \right]}
& p_z\text{-wave},\\
{\displaystyle i({\Delta}/{2})
 \hat{\sigma}_z
\left[\delta_{{\bf r},{\bf r}'+{\bf y}} - 
\delta_{{\bf r},{\bf r}'-{\bf y}} \right]}
 & p_y\text{-wave},\\
{\displaystyle ({\Delta}/{2}) 
\hat{\sigma}_z
\left[i\delta_{{\bf r},{\bf r}'+{\bf z}} -i  
\delta_{{\bf r},{\bf r}'-{\bf z}} 
-\delta_{{\bf r},{\bf r}'+{\bf y}}
+\delta_{{\bf r},{\bf r}'-{\bf y}}\right]}
& p_z+ip_y\text{-wave}
\end{cases}\,.  \label{eq:pairingsymmetry}
\eeq
\end{widetext}
The pairing potentials for the TS assume that ${\bf d}$ is directed
  along the $x$ axis in spin space.

The charge and the spin densities in the $n$-th column along the $z$ direction
are defined by 
\begin{align}
\rho(n) =& e\sum_{m=1}^{M} \sum_{\alpha}\psi_{\alpha}^\dagger(\boldsymbol{r})
\psi_{\alpha}({\bf r})\,,\\
{\bf s}(n)=&\frac{\hbar}{2}\sum_{m=1}^{M}
\sum_{\alpha,\beta}\psi_{\alpha}^\dagger({\bf r}) \hat{\boldsymbol{\sigma}}_{\alpha,\beta} 
\psi_{\beta}({\bf r})\,.
\end{align}
From the equations of motion, 
\begin{align}
\partial_t \rho(n) =&\frac{i}{\hbar}[ H, \rho(n)], \\
\partial_t {\bf s}(n) =&\frac{i}{\hbar}[ H, {\bf s}(n)], 
\end{align}
we derive the current conservation laws
\beqarray
\partial_t \rho(n)& = &- I_{ce}(n) + I_{ce}(n-1) - S_{cd}(n)\,,\label{nt}\\
\partial_t {\bf s}(n)& = &- {\bf I}_{se}(n) + {\bf I}_{se}(n-1) - 
{\bf S}_{sd}(n) - {\bf S}_v(n)\,.\label{st}
\eeqarray
On the right hand side of these equations we have terms that can be
interpreted as the divergence of a current. The first of these terms are the
familiar kinetic currents which originate
from the commutator of the densities with the hopping Hamiltonian:
\beqarray
I_{ce}(n)&=&\frac{iet}{\hbar} \sum_{m=1}^{M} \sum_{\alpha}
\left[ \psi_{\alpha}^\dagger({\bf r}+{\bf z})
\psi_{\alpha}({\bf r}) \right.\notag \\
&&\left.  - \psi_{\alpha}^\dagger({\bf r})
\psi_{\alpha}({\bf r}+{\bf z}) \right]\,,\\
{\bf I}_{se}(n)&=&\frac{it}{2} \sum_{m=1}^{M}  \sum_{\alpha,\beta}
\left[ \psi_{\alpha}^\dagger({\bf r}+{\bf z})\hat{{\bf \sigma}}_{\alpha,\beta} 
\psi_{\beta}({\bf r}) \right. \notag \\
&&\left.- \psi_{\alpha}^\dagger({\bf r})
\hat{{\bf \sigma}}_{\alpha,\beta} 
\psi_{\beta}({\bf r}+{\bf z}) \right]\,.
\eeqarray
The remaining terms in~\eq{nt} and~\eq{st} are the source terms. Specifically,
$S_{cd}(n)$ is the source term for the electric current due to  
the pair potential, ${\bf S}_{sd}(n)$ is the source term for the 
spin current due to the pair potential, and ${\bf S}_v(n)$ is the
source term for the  spin current due to the exchange potential.
They are represented by
\beqarray
S_{cd}(n)&=&\frac{-ie}{\hbar} \sum_{m=1}^{M}  \sum_{\alpha,\beta} \sum_{{\bf r}'}
\left[ \psi_{\alpha}^\dagger({\bf r}) \psi_{\beta}^\dagger({\bf r}') 
\hat{\Delta}_{\beta,\alpha}({\bf r}',{\bf r})
\right. \notag \\
&& \left. +\psi_{\alpha}({\bf r}) \psi_{\beta}({\bf r}') 
\hat{\Delta}^\ast_{\beta,\alpha}({\bf r}',{\bf r})\right]\,,\\
{\bf S}_{sd}(n)&=&\frac{-i}{2} \sum_{m=1}^{M}  \sum_{\alpha,\beta,\lambda} 
\sum_{{\bf r}'}
\left[ \psi_{\alpha}^\dagger({\bf r}) \psi_{\lambda}^\dagger({\bf r}') 
\hat{\Delta}_{\lambda,\beta}({\bf r}',{\bf r})
\boldsymbol{\sigma}^\ast_{\beta,\alpha} \right. \notag \\
&& \left.+\psi_{\alpha}({\bf r}) \psi_{\lambda}({\bf r}') 
\hat{\Delta}^\ast_{\lambda,\beta}({\bf r}',{\bf r})
\boldsymbol{\sigma}_{\beta,\alpha}\right]\,,\\
{\bf S}_v(n)&=&- \sum_{m=1}^{M}  \sum_{\alpha,\beta} 
\left[ \psi_{\alpha}^\dagger({\bf r}) \left\{ {\bf U}_M({\bf r})\times
\boldsymbol{\sigma} \right\}_{\alpha,\beta}
\psi_{\beta}({\bf r}) 
\right]\,.
\eeqarray
The magnetic source term ${\bf S}_v(n)$ has a straightforward physical
interpretation as the torque exerted by the fixed magnetic potential in the FM
layers. The pairing source terms, $S_{cd}(n)$ and ${\bf S}_{sd}(n)$, have a
more subtle origin: they account for the discrepancy between the fixed
pairing potentials~\eq{eq:pairingsymmetry} in the superconducting regions, and
the value of these pairing potentials under a self-consistent mean-field
treatment.~\cite{FurTsu1991} This discrepancy acts like a source or sink of
Cooper pairs, which must be accounted for when calculating the current. The
expectation values of these terms hence vanish under a self-consistent
treatment. Currents due to the source terms in the TS are defined 
\beqarray
I_{cd}(n)& = & -\sum_{ n+1 \leq i \leq n_0} S_{cd}(i)\,,\\
{\bf I}_{sd}(n)&= & -\sum_{n+1 \leq i \leq n_0} {\bf S}_{ds}(i)\,,\\
{\bf I}_v(n)&= & -\sum_{n+1 \leq i \leq L} {\bf S}_v(i)\,,
\eeqarray
where $n_0$ should be in the ferromagnetic layer.
We are therefore able to re-write the continuity equations~\eq{nt} and~\eq{st}
as
\beqarray
\partial_t \rho(n) & = & - {I}_c(n) + {I}_c(n-1)\,,\label{nt2}\\
\partial_t {\bf s}(n) & = & - {\bf I}^{\rm total}_{s}(n) + {\bf
  I}^{\rm total}_s(n-1)\,, \label{st2}
\eeqarray
where
\beqarray
{I}_c(n) &=& {I}_{ce}(n) + I_{cd}(n)\,,\label{ic_tb}\\ 
{\bf I}_s^{\rm total}(n) &=& {\bf I}_s(n) + {\bf I}_v(n)\,,\\
{\bf I}_s(n) &=& {\bf I}_{se}(n)+{\bf I}_{sd}(n)\,.\label{is_tb}
\eeqarray

The averages of the currents are expressed in terms of 
the Matsubara Green's function $\check{G}$ defined by
\beqarray
\check{G}({\bf r},{\bf r}',\tau-\tau')
&=& - \left\langle T_\tau
\Psi^{}({\bf r})\Psi^\dagger({\bf r}')
\right\rangle \notag \\
&= &T\sum_{\omega_n} \check{G}({\bf r},{\bf r}',\omega_n)
e^{-i\omega_n(\tau-\tau')}\,,
\eeqarray
where $\omega_n=(2n+1)\pi T$ are the Matsubara frequencies at temperature
$T$. Specifically, we write\cite{asanospin}
\begin{widetext}
\beqarray
I_{ce}(n)&=&\frac{-iet}{2\hbar} \sum_{m=1}^M T\sum_{\omega_n}
\textrm{Tr} \left[
\check{G}({\bf r}+{\bf z},{\bf r},\omega_n)
-\check{G}({\bf r},{\bf r}+{\bf z},\omega_n)\right]\,,\\
S_{ce}(n)&=&\frac{ie}{\hbar} \sum_{m=1}^M T\sum_{\omega_n} \sum_{{\bf r}'}
\textrm{Tr} \left[
\check{G}({\bf r},{\bf r}',\omega_n)
\left(\begin{array}{cc}
0 & {\bf \Delta}({\bf r}',{\bf r})\\
{\bf \Delta}^\ast({\bf r}',{\bf r}) & 0
\end{array}\right)
\right]\,,\\
{\bf I}_{sq}(n) &=&\frac{-it}{4} \sum_{m=1}^M T\sum_{\omega_n}
\textrm{Tr} \left[\left(
\check{G}({\bf r}+{\bf z},{\bf r},\omega_n)
-\check{G}({\bf r},{\bf r}+{\bf z},\omega_n)\right)
\left(\begin{array}{cc}
\boldsymbol{\sigma} & 0\\
0& \boldsymbol{\sigma}^\ast
\end{array}\right)
\right]\,,\\
{\bf S}_{sd}(n)&=&\frac{i}{2} \sum_{m=1}^M T\sum_{\omega_n} \sum_{{\bf r}'}
\textrm{Tr} \left[
\check{G}({\bf r},{\bf r}',\omega_n)
\left(\begin{array}{cc}
0 & {\bf \Delta}({\bf r}',{\bf r})\\
{\bf \Delta}^\ast({\bf r}',{\bf r}) & 0
\end{array}\right)
\left(\begin{array}{cc}
\boldsymbol{\sigma} & 0\\
0& \boldsymbol{\sigma}^\ast
\end{array}\right)
\right]\,,\\
{\bf S}_{v}(n)&=&\frac{-1}{2} \sum_{m=1}^M T\sum_{\omega_n} 
\textrm{Tr} \left[
\check{G}({\bf r},{\bf r},\omega_n)
\left(\begin{array}{cc}
{\bf U}_M({\bf r}) \times \boldsymbol{\sigma} &  0\\
0& -{\bf U}_M({\bf r}) \times \boldsymbol{\sigma}^\ast
\end{array}\right)
\right]\,.
\eeqarray
\end{widetext}
The recursive Green's function method enables us to numerically calculate
the Green's function, and hence evaluate the above
equations.~\cite{Lee,asano01}

The charge current ${I}_c(n)$ in~\eq{ic_tb} is independent of $n$,
as required by the charge conservation law. Spin must also be conserved, and
so we find that ${\bf I}_s^{\rm total}(n)$ is vanishing for all $n$ because
the spin current cannot flow in the SS. This result contradicts the prediction
of~\Sec{sec:perturbation} that there is a  Josephson spin current
in the TS. The paradox can be resolved by noting that neither the perturbation
theory nor the Green's function method can properly account for the transfer
of spin to the ferromagnetic barrier, as in both theories the magnetic moment
is assumed fixed by the constant exchange potential ${\bf U}_M({\bf
  r})$. Rigorously accounting for the conservation of spin in such a situation
naturally leads to the conclusion of vanishing spin current. It is
nevertheless reasonable to identify the Josephson spin current with the
current ${\bf I}_s(n)$, which produces a torque on the
ferromagnetic barrier, and to hence regard ${\bf I}_v(n)$ as a compensating
current necessary to maintain the constant exchange potential.
Although ${\bf I}_s(n)$ depends on $n$ in the ferromagnet, it is
independent of $n$ in the TS. In the following we only consider this
spin current in the TS, in order to make contact with the
perturbation theory.

In the following we present results for a junction of width $M=10$ and
ferromagnetic barrier length $L=10$. In units of the transfer integral $t$
we take $|{\bf U}_M|=0.1$ for the exchange potential, and $\mu=2$ for the
chemical potential. The pairing potential has weak-coupling
temperature-dependence, with zero-temperature magnitude $\Delta_0=0.01$. 
We have confirmed that the transport properties are qualitatively insensitive
to choices of these parameters.  
Since ${\bf U}_M\times {\bf d} \parallel {\bf z}$ as shown in  
Fig.~\ref{asano1}, the $x$ and $y$ components of the spin current are zero.
To compare with the analytical predictions of~\Sec{sec:perturbation}, we fix
the temperature at $T=0.5T_c$ in the tunneling regime.

\subsection{$s$-wave singlet superconductor}

\begin{figure}
\includegraphics[width=0.95\columnwidth,clip=true]{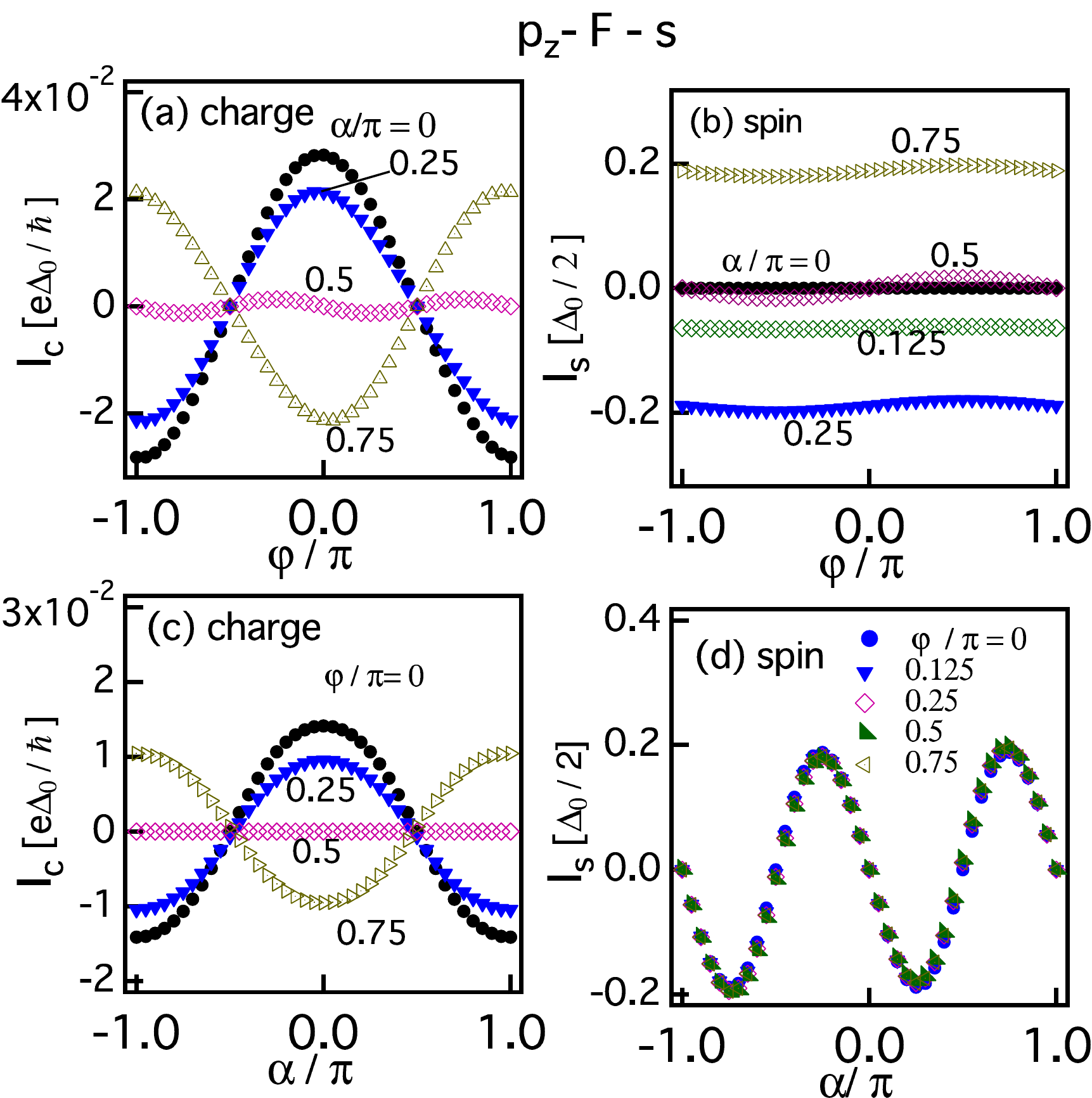}
\caption{(color online) Josephson currents in the $p_z$-$F$-$s$ junction. (a)
  Charge and (b) 
  $z$-component spin currents as a function of $\phi$ for fixed $\alpha$. (c)
  Charge and (d) $z$-component spin currents as a function of $\alpha$ for
  fixed $\phi$. We choose the parameters as $M=L=10$, $\mu=2t$, $|{\bf
    U}_M|=0.1t$, $\Delta_0=0.01t$, and $T=0.5T_c$.}
\label{fig_s_px}
\end{figure}

\begin{figure}
\includegraphics[clip,width=0.95\columnwidth,clip=true]{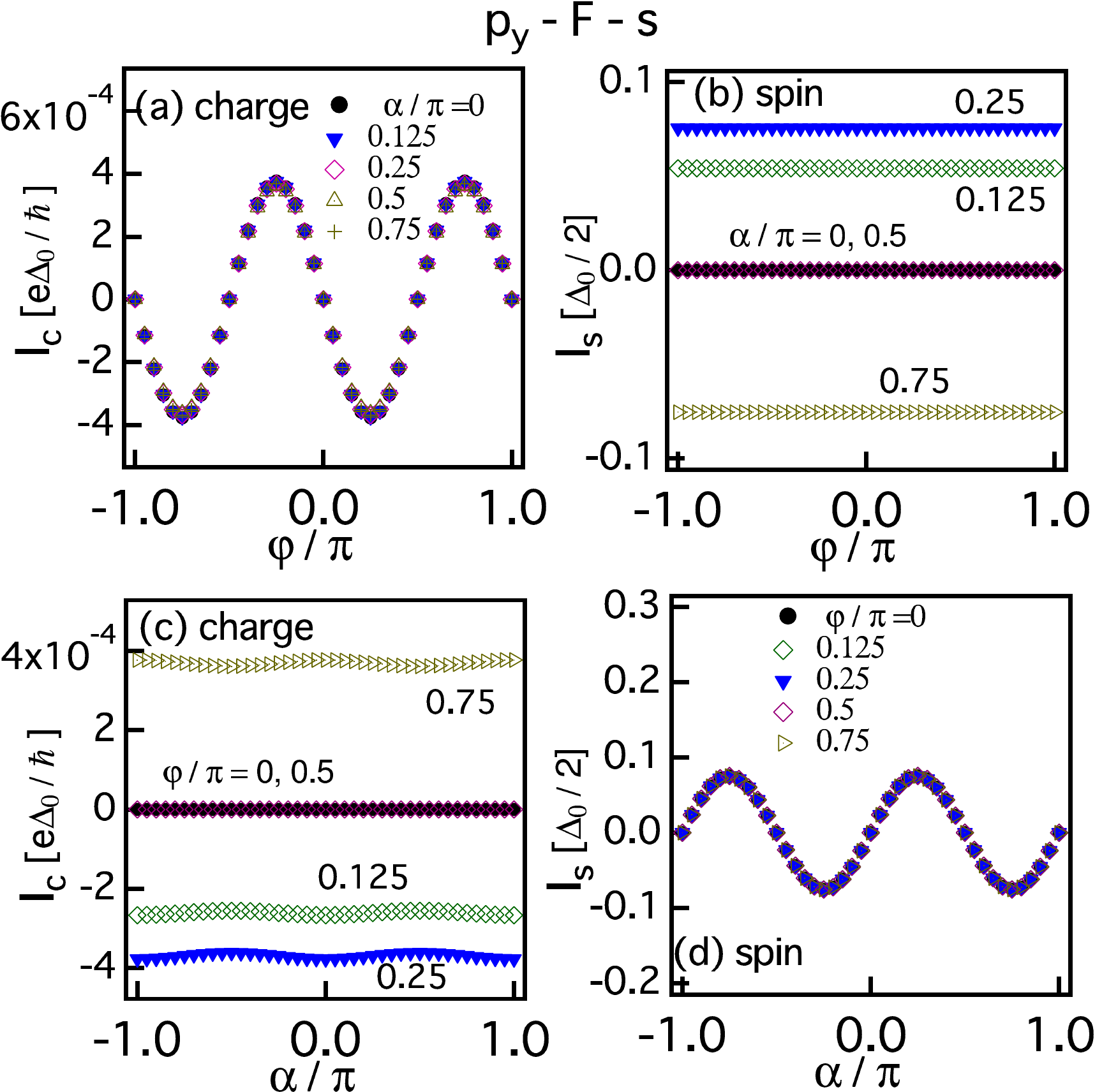}
\caption{(color online) Josephson currents in the $p_y$-$F$-$s$ junction. (a) Charge and (b)
  $z$-component spin currents as a function of $\phi$ for fixed $\alpha$. (c)
  Charge and (d) $z$-component spin currents as a function of $\alpha$ for
  fixed $\phi$.
  The parameter values are fixed as in Fig.~\ref{fig_s_px}. }
\label{fig_s_py}
\end{figure}

In this subsection we present results for the Josephson currents in a TFS
junction between an $s$-wave SS and each of the three
different TS states listed
in~\eq{eq:pairingsymmetry}. Commencing 
with the $p_z$-wave TS (the $p_z$-$F$-$s$ junction), in Fig.~\ref{fig_s_px} we
plot the charge and spin 
currents as functions of the phase $\phi$ and angle $\alpha$. As
can be seen 
in panels (a) and (c), the dominant term in the charge current is
\beq
I_c = \widetilde{I}_c\cos(\phi)\cos(\alpha)\,, \label{cspx}
\eeq
with a much weaker contribution $\propto\sin(2\phi)$ visible at
$\alpha=0.5\pi$ in panel (a). This is clearly consistent with the perturbation
theory predictions. The spin current also agrees with the
perturbative analysis, with the numerical results well described by 
\beq
I_{s,z} = \widetilde{I}_s\sin(2\alpha) + \widetilde{I}^\prime_s\sin(\phi)\sin(\alpha)\,,\label{sspx}
\eeq
where $\widetilde{I}_s \gg \widetilde{I}^\prime_s$. Indeed, in panel (d) the
spin current vs $\alpha$ curves at different $\phi$ almost overlap due to the
very weak $\phi$-dependence. The much smaller coefficient of the
$\phi$-dependent term was anticipated in our tunneling Hamiltonian
  analysis.

\begin{figure}
\includegraphics[width=0.95\columnwidth,clip=true]{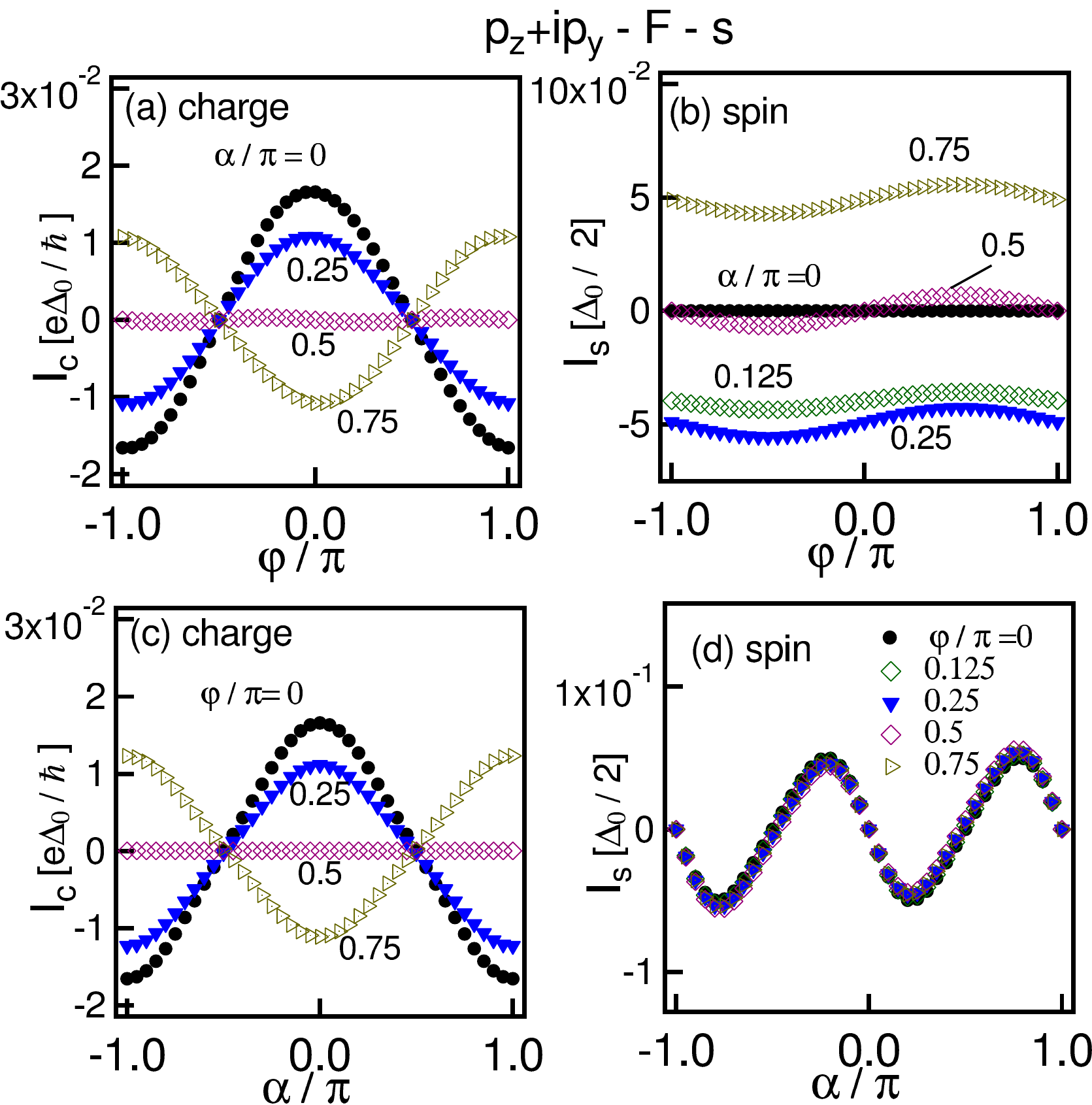}
\caption{(color online) Josephson currents in the $(p_z+ip_y)$-$F$-$s$ junction. (a) Charge and (b)
  $z$-component spin currents as a function of $\phi$ for fixed $\alpha$. (c)
  Charge and (d) $z$-component spin currents as a function of $\alpha$ for
  fixed $\phi$. The parameter values are fixed as in Fig.~\ref{fig_s_px}.}
\label{fig_s_chiralp}
\end{figure}

We now consider the results for the $p_y$-wave TS, which are shown
in~Fig.~\ref{fig_s_py}. In contrast to the $p_z$-wave TS, the dominant
contribution to the charge current is $\propto\sin(2\phi)$, there is only
very weak dependence upon $\alpha$, and the maximum critical
current is much smaller. Since the $\propto\sin(2\phi)$ term originates
  from coherent tunneling of two Cooper pairs,~\cite{PalHaeMaa1977} these
  results are consistent with our prediction of vanishing charge current due
  to single-Cooper-pair tunneling processes. For 
the spin current we find $I_{s,z} = 
\widetilde{I}_s\sin(2\alpha)$ to excellent approximation, in perfect agreement
with the perturbation theory predictions.
Note that the spin current has opposite sign compared to the $p_z$-wave
junction as expected.

In~\fig{fig_s_chiralp} we present the currents for the ($p_z+ip_y$)-wave TS
state. As predicted in~\Sec{sec:perturbation}, the results for this junction
are very similar to those for the $p_z$-wave TS, as the lowest-order
Josephson coupling proceeds through the $p_z$-component
of the chiral $p$-wave gap. The
results are therefore summarized by  Eqs.~(\ref{cspx}) and (\ref{sspx}).

\subsection{$d_{yz}$-wave singlet superconductor }

\begin{figure}
\includegraphics[width=0.95\columnwidth,clip=true]{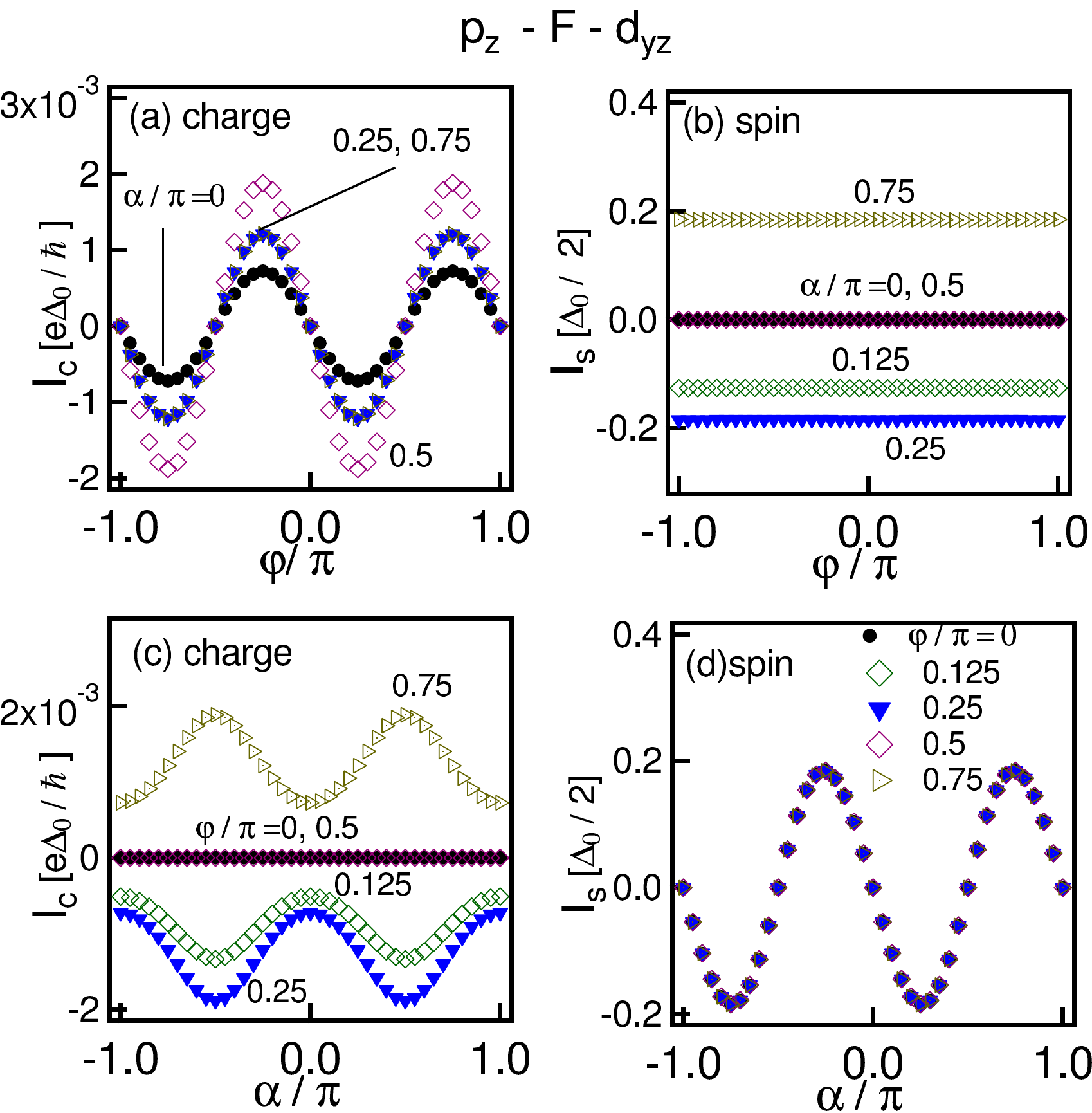}
\caption{(color online) Josephson currents in the $p_z$-$F$-$d_{yz}$ junction. (a) Charge and (b)
  $z$-component spin currents as a function of $\phi$ for fixed $\alpha$. (c)
  Charge and (d) $z$-component spin currents as a function of $\alpha$ for
  fixed $\phi$. The parameter values are fixed as in Fig.~\ref{fig_s_px}.}
\label{fig_dxy_px}
\end{figure}

The parity requirement for the superconducting gaps leads us to expect
qualitatively different behaviour upon replacing the $s$-wave superconductor
by a $d_{yz}$-wave superconductor due to the even and odd dependence on $k_y$,
respectively. To test this, we 
repeat the above analysis for a $d_{yz}$-wave pairing symmetry in the
SS. Starting with the $p_z$-wave TS, in~\fig{fig_dxy_px} we find  
that the charge current is approximately given by
\beq
I_c = \widetilde{I}_c\sin(2\phi) + \widetilde{I}^\prime_c\sin(2\phi)\cos(2\alpha)\,, \label{cdxypx}
\eeq
with $\widetilde{I}_c$ and $\widetilde{I}^\prime_c$ of comparable
magnitude. This is clearly consistent with the predicted absence of
single-Cooper-pair tunneling processes when only one of the order
parameters is odd in $k_y$. The spin current is independent of $\phi$ and
has the approximate form $I_{S,z} = \widetilde{I}_s\sin(2\alpha)$,
characteristic of the contribution due to spin-flip reflection.

\begin{figure}
\includegraphics[width=0.95\columnwidth,clip=true]{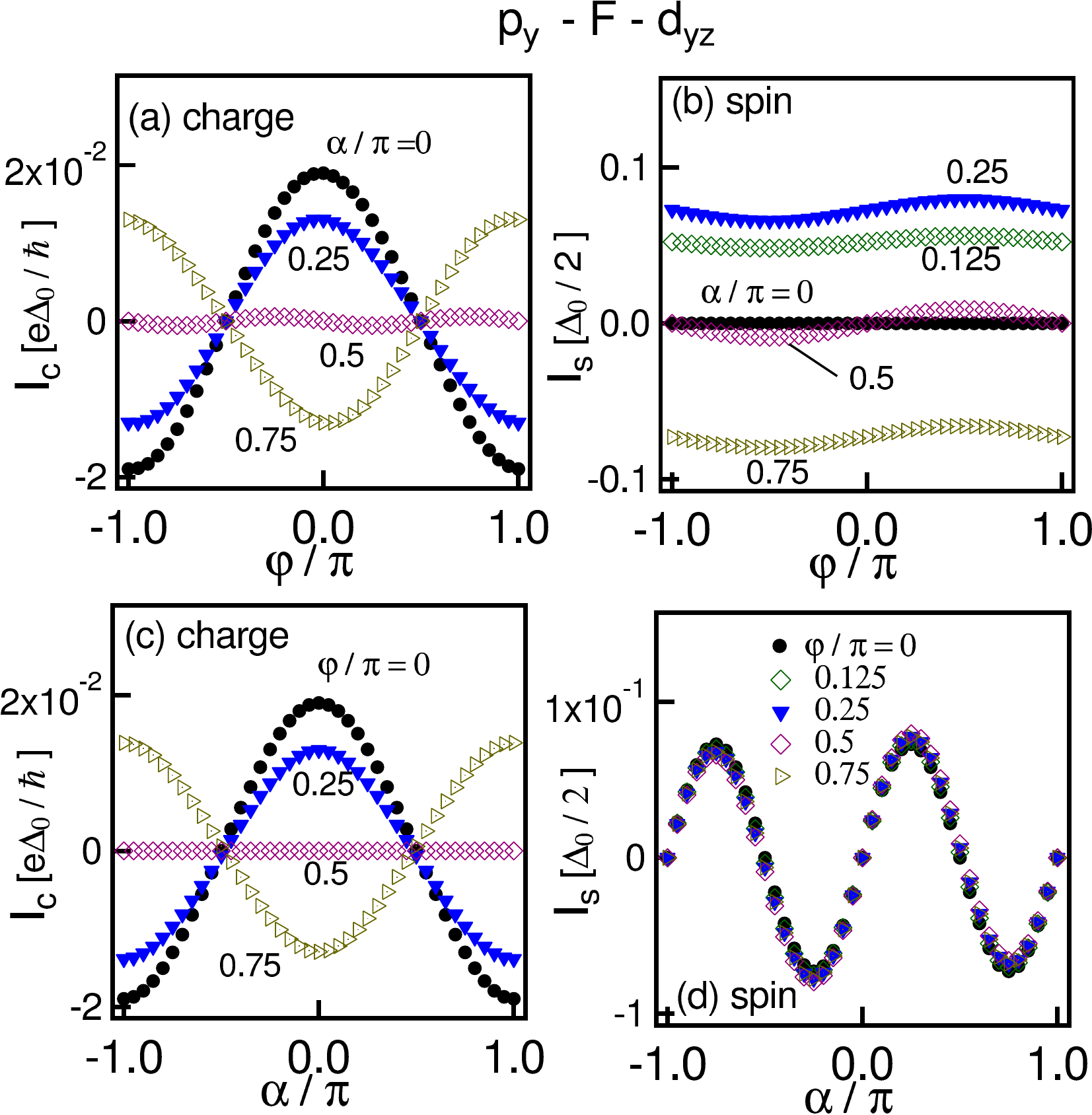}
\caption{(color online) Josephson currents in the $p_y$-$F$-$d_{yz}$ junction. (a) Charge and (b)
  $z$-component spin currents as a function of $\phi$ for fixed $\alpha$. (c)
  Charge and (d) $z$-component spin currents as a function of $\alpha$ for
  fixed $\phi$. The parameter values are fixed as in Fig.~\ref{fig_s_px}.}
\label{fig_dxy_py}
\end{figure}

\begin{figure}[t!]
\includegraphics[width=0.95\columnwidth,clip=true]{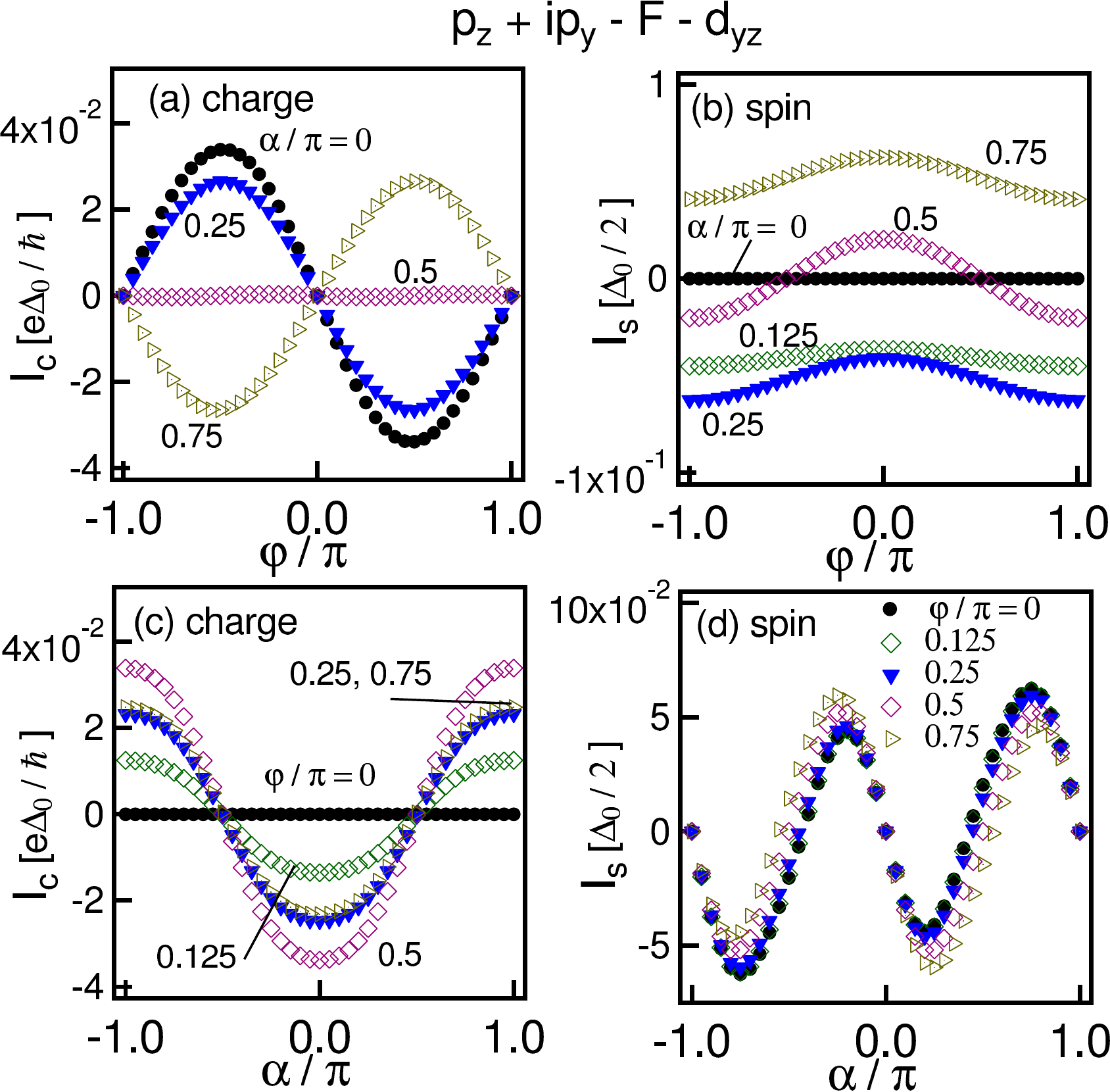}
\caption{(color online) Josephson currents in the $(p_z+ip_y)$-$F$-$d_{yz}$ junction. (a) Charge and (b)
  $z$-component spin currents as a function of $\phi$ for fixed $\alpha$. (c)
  Charge and (d) $z$-component spin currents as a function of $\alpha$ for
  fixed $\phi$. The parameter values are fixed as in Fig.~\ref{fig_s_px}.}
\label{fig_dxy_chiralp}
\end{figure}

In Fig.~\ref{fig_dxy_py} we show the charge and spin currents for the
$p_y$-wave TS pairing symmetry. Since both the singlet and triplet gaps are
odd in $k_y$, we expect that a charge current is realized at the lowest order
of perturbation theory. Indeed, as can be seen in panels (a) and (c), the
numerical results for the charge current are well approximated by~\eq{cspx},
and the charge current is an order of magnitude larger than for the $p_z$-wave
junction. As expected, there is also a $\phi$-dependent contribution to the
spin current, which is again  of the form~\eq{sspx}.

To conclude our survey, in Fig.~\ref{fig_dxy_chiralp} we plot the charge and
spin 
currents for the chiral $p$-wave TS state. In this junction the $d_{yz}$
symmetry couples to $i p_y$ component of the ($p_z+ip_y$)-wave symmetry, and
so the factor of $i$ gives an additional $\pi/2$ phase shift.
Therefore, the lowest-order currents in this junction should be approximately
obtained from the 
pure $p_y$-wave case with the replacement $\phi\rightarrow\phi + \pi/2$.
Indeed the numerical results in Fig.~\ref{fig_dxy_chiralp} are consistent with
the relations
\beqarray
I_c &=& \widetilde{I}_c\sin(\phi)\cos(\alpha)\,, \label{cdxychiralp}\\
I_{s,z} &=& \widetilde{I}_s\sin(2\alpha) + \widetilde{I}^\prime_s\cos(\phi)\sin(\alpha)\,. \label{sdxychiralp}
\eeqarray
This agrees with the predicted currents in~\tab{tab:lowestorder}. 

In contrast to the charge current, the spin current hardly changes 
upon replacing the $s$-wave SS by the $d_{xy}$-wave SS. This reflects the
dominance of reflection processes, which are insensitive to the superconductor
on the other side of the junction. 

\section{Role of Andreev bound states} \label{sec:quasiclassics}

The perturbation theory of~\Sec{sec:perturbation} gives a good
description of the transport when multiple-Cooper-pair tunneling processes
make an insignificant contribution to the current, 
i.e. when higher-order terms in 
the perturbation expansion can be neglected. Although this is always the case
sufficiently close to the transition temperature of the superconductors, 
resonant tunneling through Andreev bound states can cause
large deviations from perturbation theory predictions at zero temperature.
This is often particularly pronounced in junctions where the gap of
the superconductors is odd in the momentum component perpendicular to
the interface.~\cite{KasTan2000,KwoSenYak2004}  
It is therefore likely that in some of the junctions studied above, e.g. the
junction between a $p_z$-wave TS and an $s$-wave SS, there will be strong
contributions to the current from higher harmonics in $\phi$ and $\alpha$ at
low temperatures. In order to estimate the importance of this effect, here we
consider an analytically-tractable model of the TFS junction where the current
is due entirely to tunneling through Andreev bound states.

\subsection{One-dimensional model junction}

We study a one-dimensional continuum model of the TFS junction with $p_z$-
and $s$-wave orbital symmetries for the TS and SS, respectively. Our
analysis here closely follows that of Refs.~\onlinecite{Kastening06}
and~\onlinecite{Brydon08} for a TFT junction. The energies of the Andreev
bound states are obtained by means of solving the Bogoliubov-de Gennes
equation $H\Psi(z)=E\Psi(z)$, with Hamiltonian
\begin{widetext} 
\begin{eqnarray}
H=\left(\begin{array}{llll}
-\frac{\hbar^{2}}{2m}\frac{\partial^2}{\partial z^2}+U_{0}\delta(z)-\mu &
-e^{-i\alpha}U_M\delta(z) &
-i\Theta(-z)\Delta_{t}\frac{\partial}{\partial z} & \Theta(z)\Delta_{s}e^{-i\phi} \\
-e^{i\alpha}U_M\delta(z) &
-\frac{\hbar^{2}}{2m}\frac{\partial^2}{\partial z^2}+U_{0}\delta(z)-\mu  &
-\Theta(z)\Delta_{s}e^{-i\phi} & i\Theta(-z)\Delta_{t}\frac{\partial}{\partial z} \\
-i\Theta(-z)\Delta_{t}\frac{\partial}{\partial z} &
-\Theta(z)\Delta_{s}e^{i\phi} & \frac{\hbar^{2}}{2m}\frac{\partial^2}{\partial
  z^2}-U_{0}\delta(z)+\mu & e^{i\alpha}U_M\delta(z) \\
\Theta(z)\Delta_{s}e^{i\phi} & i\Theta(-z)\Delta_{t}\frac{\partial}{\partial z} & e^{-i\alpha}U_M\delta(z) & \frac{\hbar^{2}}{2m}\frac{\partial^2}{\partial z^2}-U_{0}\delta(z)+\mu 
\end{array}\right)\,.
\label{TFS_BdG_Hamiltonian}
\end{eqnarray}
\end{widetext}
For simplicity we assume that the effective mass $m$ and chemical potential
$\mu$ are the same on either side of the junction, and hence the Fermi
wavevectors $k_{F,\nu}$ in each superconductor are also the same,
i.e. $k_{F,\nu}=k_F$. The TS and SS are 
described by the pairing potentials $\Delta_t$ and $\Delta_s$,
respectively. The barrier is modeled as a $\delta$-function with charge
scattering potential $U_0$ and magnetic scattering potential ${\bf U_M}=
U_M(\cos\alpha\hat{\bf e}_x + \sin\alpha\hat{\bf e}_y)$.

The Andreev bound states have energy lying within the bulk gap of the two
superconductors, i.e. $|E|<\min\{k_F|\Delta_t|,|\Delta_s|\}$, and are hence
exponentially localized at the interface. Within the Andreev approximation,
where the superconducting gap is assumed negligible compared to the Fermi
energy, an appropriate ansatz for the wavefunction of these states is
\beq
\Psi_{\nu}(z) =
e^{-\nu\kappa_{\nu}z}\left(\Psi_{\nu,+}e^{+ik_Fz}+\Psi_{\nu,-}e^{-ik_Fz}\right)\Theta(\nu
z)\,.
\eeq
Following the notation of~\Sec{sec:perturbation}, $\nu=s$~($t$) denotes the
singlet (triplet) side, and as a factor $\nu = 1$ ($-1$). 
$\kappa_{\nu}$ is the inverse decay length in the $\nu$ superconductor. The
spinors $\Psi_{\nu,\pm}$ are defined
\begin{subequations} \label{eq:ansatzspinors}
\begin{eqnarray}
\Psi_{t,\pm} & = & \left(\begin{array}{cccc}a_{t,\pm}, & b_{t,\pm}, &
  \pm e^{\pm i\gamma_t}a_{t,\pm}, & \mp e^{\pm i\gamma_t}b_{t,\pm}\end{array}\right)^{T}\,, \\
\Psi_{s,\pm} & = & \left(\begin{array}{cccc}a_{s,\pm}, & b_{s,\pm}, &
  -e^{i(\phi \mp \gamma_s)}b_{s,\pm}, & e^{i(\phi \mp
    \gamma_s)}a_{s,\pm}\end{array}\right)^{T}\,,\notag \\
\end{eqnarray}
\end{subequations}
where the subscript $\left\{+,-\right\}$ indicates the direction of
propagation, and the phases $\gamma_\nu$ are given by
\begin{subequations}
\begin{gather}
\cos\gamma_t = \frac{E}{k_F|\Delta_t|}, \qquad \sin\gamma_{t}=\frac{\hbar^{2}\kappa_{t}}{m|\Delta_{t}|}\,,\\
\cos\gamma_s = \frac{E}{|\Delta_s|}, \qquad \sin\gamma_{s}=\frac{\hbar^{2}\kappa_{s}k_{F}}{m|\Delta_{s}|}\,.
\end{gather}
\end{subequations}
The $a_{\nu,\pm}$ and $b_{\nu,\pm}$ appearing in~\eq{eq:ansatzspinors} are
constants to be determined by the boundary conditions obeyed by the
wavefunction at the interface. In addition to continuity of the wavefunction
across the junction,
\beq
\Psi_{t}(z = 0^{-}) = \Psi_{s}(z = 0^{+})\,,
\eeq
we require that the derivative obeys
\beqarray
\lefteqn{\partial_{z}\Psi_{s}(z)|_{z = 0^{+}} - \partial_{z}\Psi_{t}(z)|_{z =
    0^{-}}} \notag \\
& = & 2k_F\left(\begin{array}{cccc}
Z & - ge^{-i\alpha} & 0 & 0 \\
-g e^{i\alpha} & Z & 0 & 0 \\
0 & 0 & Z & -g e^{i\alpha} \\
0 & 0 & -g e^{-i\alpha} & Z
\end{array}\right)\Psi_{s}(z=0^+)\notag\,, \\
\eeqarray
in order to conserve probability. Here we use the dimensionless
parameters    
\begin{eqnarray}
Z=\frac{k_FU_{0}}{2\mu}\;, \qquad g=\frac{k_FU_M}{2\mu}\;,
\label{BdG_dimensionless_parameters}
\end{eqnarray}
to characterize the strength of barrier potentials.
The boundary conditions give eight equations for the coefficients
$a_{\nu,\pm}$ and $b_{\nu,\pm}$, and have nontrivial solution when the energy
of the bound state satisfies the equation
\begin{eqnarray}
0 &= & 2+8g^{2}+8g^{2}\cos(2\alpha)-2\cos(2\phi)-W\cos^{2}(\gamma_{t})
\nonumber \\
&&-\left[4+8g^{2}+8g^{2}\cos(2\alpha)\right]\cos^{2}(\gamma_{s})
\nonumber \\
&&+\left[4+W+16g^{2}\right]
\cos^{2}(\gamma_{t})\cos^{2}(\gamma_{s})
\nonumber \\
&&-2\left[2g^{2}+2Z^{2} + 1\right]
\sin(2\gamma_{t})\sin(2\gamma_{s})
\nonumber \\
&&+16g\cos(\alpha)\sin(\phi)\cos(\gamma_{t})\cos(\gamma_{s})
\nonumber \\
&&-16g\cos(\alpha)\sin(\phi)\sin(\gamma_{t})\sin(\gamma_{s})\,,
\label{TFS_energy_condition}
\end{eqnarray}
where 
\beq
W=4\left(2Z^{2}+1\right) +16\left(g^{2}- Z^{2}\right)^2\;.
\label{BdG_energy_condition}
\eeq

\subsection{Analytical solution at $k_{F}\Delta_{t}=\Delta_{s}$}

The parameters that characterize triplet and singlet superconductors are, of
course, independent from each other, as the two superconductors cannot be made
from the same material. Therefore there are numerous choices to realize the
parameters in Eq. (\ref{BdG_dimensionless_parameters}), according to the
materials from which the junction is made. Nevertheless, in the
limit when the pairing potentials of the singlet and triplet
superconductor are the same, i.e., $k_{F}\Delta_{t}=\Delta_{s}$, it
is possible to express the Josephson charge and spin currents entirely in
terms of the Andreev bound state energies. Although this is a highly-idealized
situation, it clearly reveals the influence of resonant tunneling on the
currents, which we expect to remain qualitatively valid for a more realistic 
model of the junction.

\begin{figure}
\centering
\includegraphics[width=\columnwidth,clip=true]{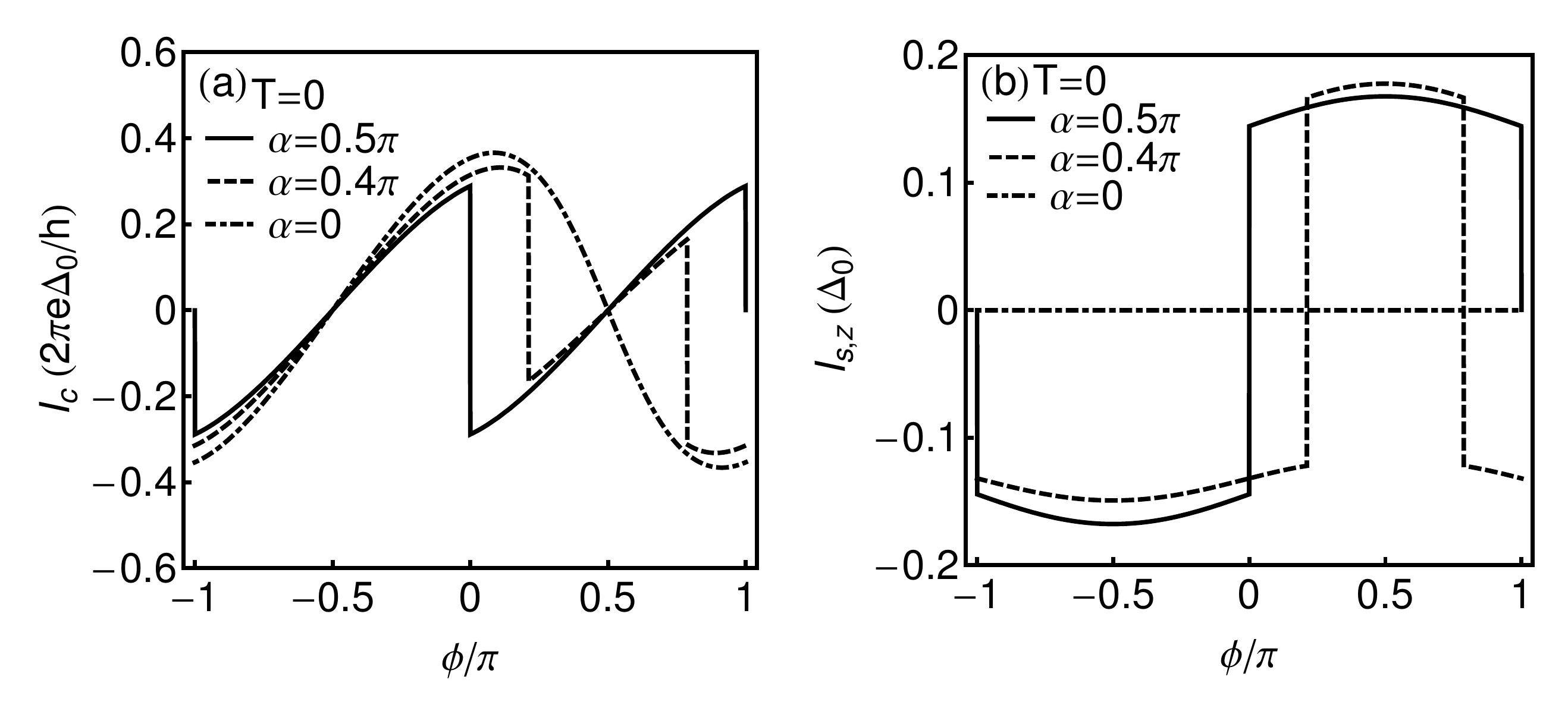} \\
\includegraphics[width=\columnwidth,clip=true]{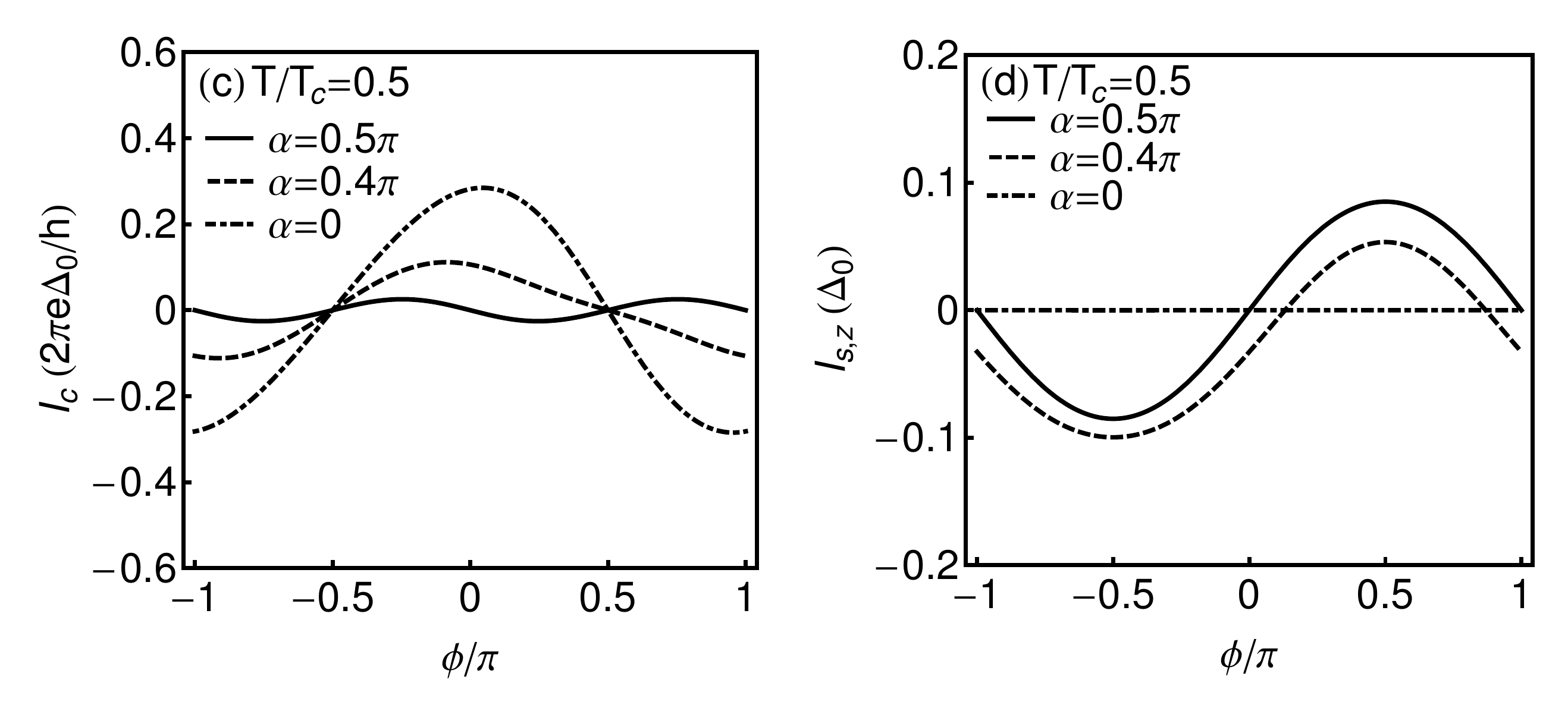}
\caption{(a) Josephson charge and (b) spin current for the case $g=1$,
  $Z=0$, at zero temperature $T=0$. (c) Josephson charge and (d) spin current
  for the same parameter set at finite temperature $T/T_{c}=0.5$.} 
\label{fig:charge_spin_current}
\end{figure}

When $k_{F}\Delta_{t}=\Delta_{s}$, the bound state energy parameters are equal
$\gamma_{L}=\gamma_{R}$, and Eq. (\ref{TFS_energy_condition}) is drastically
simplified 
\begin{eqnarray}
\frac{1}{D^{2}}\left(\frac{E}{|\Delta_{s}|}\right)^{4}-4A\left(\frac{E}{|\Delta_{s}|}\right)^{2}+4B^{2}=0\,, 
\label{E4_E2_C}
\end{eqnarray}
where
\begin{subequations}
\begin{eqnarray}
D&=&\left[g^{4}+2g^{2}\left(1-Z^{2}\right)+\left(1+Z^{2}\right)^{2}\right]^{-1/2}\,,
\\
A&=&\frac{1}{4}\left[\frac{1}{D^{2}}-g^{2}\sin^{2}(\alpha)
-2g\cos(\alpha)\sin(\phi)\right]\,,
 \\
B&=&\frac{g}{2}\cos(\alpha)-\frac{1}{4}\sin(\phi)\,.
\end{eqnarray}
\end{subequations}
The positive Andreev bound state energies are hence found to be 
\beq
\frac{E_{a,b}}{|\Delta_{s}|}=\sqrt{D}|\sqrt{DA+B}\pm \sqrt{DA-B}|\,. \label{bound_states}
\eeq
The Josephson charge and $z$-spin currents are defined~\cite{currentformula,TSFM}
\begin{subequations} \label{charge_spin_current}
\begin{eqnarray}
I_{c}&=&-\frac{e}{\hbar}\sum_{l=a,b}\frac{\partial
  E_{l}}{\partial\phi}\tanh\frac{E_{l}}{2k_{B}T}\,, \\
I_{s,z}&=&\frac{1}{4}\sum_{l=a,b}\frac{\partial E_{l}}{\partial \alpha}\tanh\frac{E_{l}}{2k_{B}T}\,.
\end{eqnarray}
\end{subequations}
Inserting~\eq{bound_states}
into~\eq{charge_spin_current}  we hence obtain
\begin{subequations} \label{charge_spin_current_specific}
\begin{eqnarray}
I_c & = &
\frac{e|\Delta_{s}|}{8\hbar}\sum_{\sigma=\pm}\sqrt{\frac{D}{DA+\sigma
    B}} \nonumber \\
&& \times \left[2Dg\cos(\alpha)\cos(\phi) +\sigma\cos(\phi)\right]
\nonumber \\
&&\times \left(\tanh\frac{E_{a}}{2k_{B}T} +
\sigma\text{sign}(B)\tanh\frac{E_{b}}{2k_{B}T}\right)\,, \\
I_{s,z} &=&
-\frac{|\Delta_{s}|}{16}\sum_{\sigma=\pm}\sqrt{\frac{D}{DA+\sigma
    B}} \nonumber \\
&& \times \left[\frac{D}{2}g^{2}\sin(2\alpha)
  -Dg\sin(\alpha)\sin(\phi)+ \sigma g\sin(\alpha)\right]
\nonumber \\
&&\times \left(\tanh\frac{E_{a}}{2k_{B}T} +
\sigma\text{sign}(B)\tanh\frac{E_{b}}{2k_{B}T}\right) \,.
\end{eqnarray}
\end{subequations}
We plot the Josephson currents at zero and finite temperature in
the upper and lower panels of~\fig{fig:charge_spin_current},
respectively. To obtain the finite temperature results we assume that
both gaps display BCS weak-coupling temperature-dependence  with 
critical temperature $T_c$. We find
that the barrier potential $Z$ mainly affects the amplitude  but not
the form of the current-phase relation, so hereafter we only present
results for $Z=0$.

The zero temperature results for the charge and spin currents at
$\alpha=0$ are in good agreement with the perturbation theory
predictions. Rotating the magnetic moment towards the $y$-$z$ plane,
however, we observe the current-phase relations display sharp jumps
at the zero-energy crossings of the Andreev bound
states. The Fourier decomposition with respect to the phase difference
hence contains a large contribution from higher harmonics, and thus
multiple Cooper pair tunneling processes are important in the
zero-temperature limit. Indeed, we see
in~\fig{fig:charge_spin_current}(a) that the maximum current at
$\alpha=0.5\pi$ is comparable to that at 
$\alpha=0$, whereas in the tunneling regime we expect that the latter
should be much larger than the former. 
In contrast, the currents at half the critical temperature [panels (c) and
  (d)] are in much better agreement with the perturbation theory predictions
and the lattice model calculations. In particular, the amplitude of the
charge current-phase relation at $\alpha=0.5\pi$ is now much smaller than that
at $\alpha=0$.

The greatest deviations between the exact and perturbative results for the
current at zero temperature occurs for angles near
to $\alpha=\pm0.5\pi$, 
where  we find jump discontinuities in the current due to zero-energy
crossings of the Andreev bound states which are present for
$||\alpha| - 0.5\pi| \leq \arcsin(1/2g)$. For $g\gg 1$, the
zero-temperature current is well approximated by
\beq
I_{c}=  \frac{e\Delta_0}{2\hbar}\frac{1}{g^2}\mbox{sign}(\cos(\alpha))\cos(\phi)\,.
\eeq
Note that the amplitude of the current goes as $g^{-2}$, instead of $g^{-3}$ as
predicted in~\eq{eq:pt:IJ}: enhancement of the low-temperature current above
the perturbation theory predictions is a  
well-known consequence of the presence of zero-energy Andreev
states.~\cite{KwoSenYak2004} We now turn to the spin current: In the
  large-$g$ 
limit the phase-dependent component becomes negligible, and
we find  
\beq
I_{s,z}= -\frac{\Delta_0}{4}\frac{1}{g}\mbox{sign}(\cos(\alpha))\sin(\alpha)\,.
\eeq
Although the current is again enhanced beyond the perturbation theory
predictions, the two approaches agree that reflection processes dominate in the
limit of a strong magnetic barrier.

\begin{figure}
\centering
\includegraphics[width=\columnwidth,clip=true]{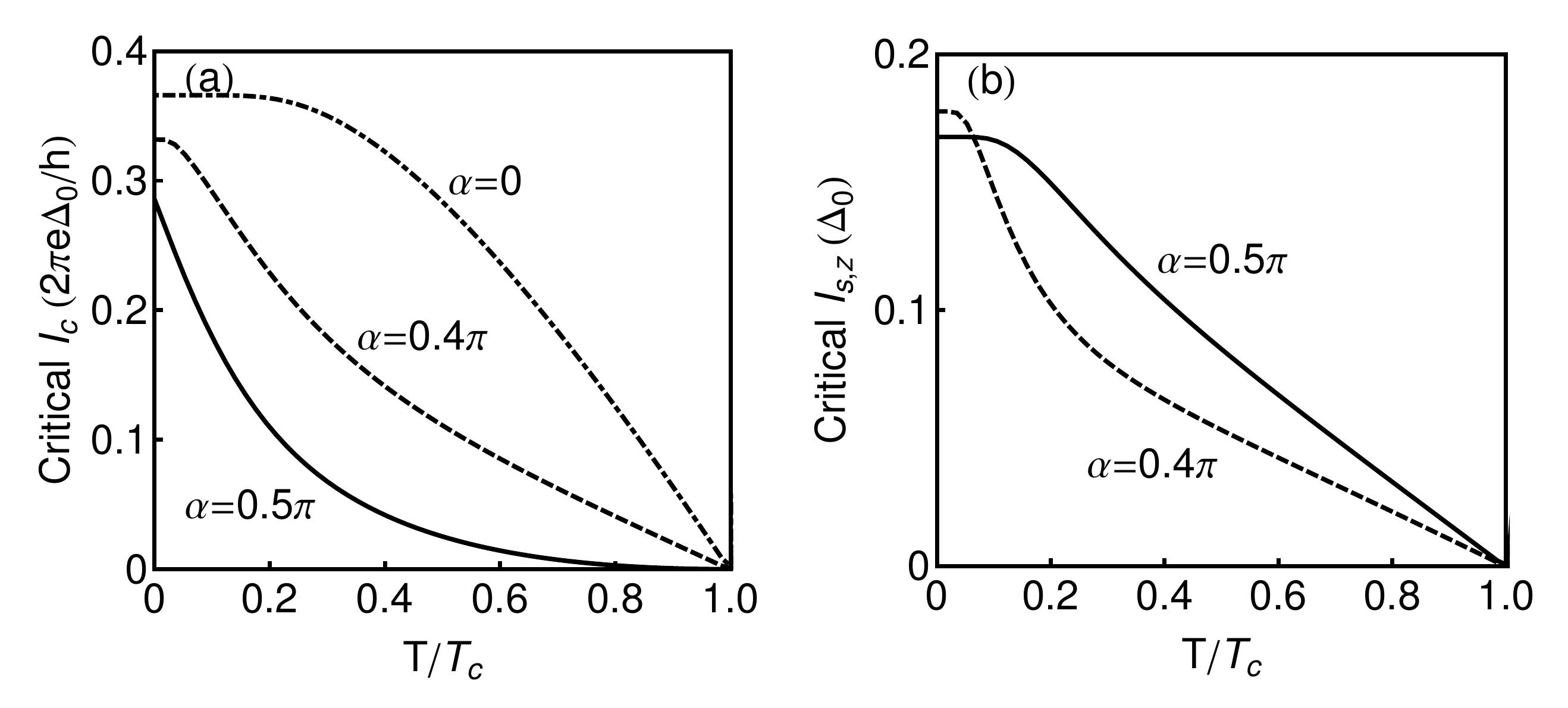} \\
\includegraphics[width=\columnwidth,clip=true]{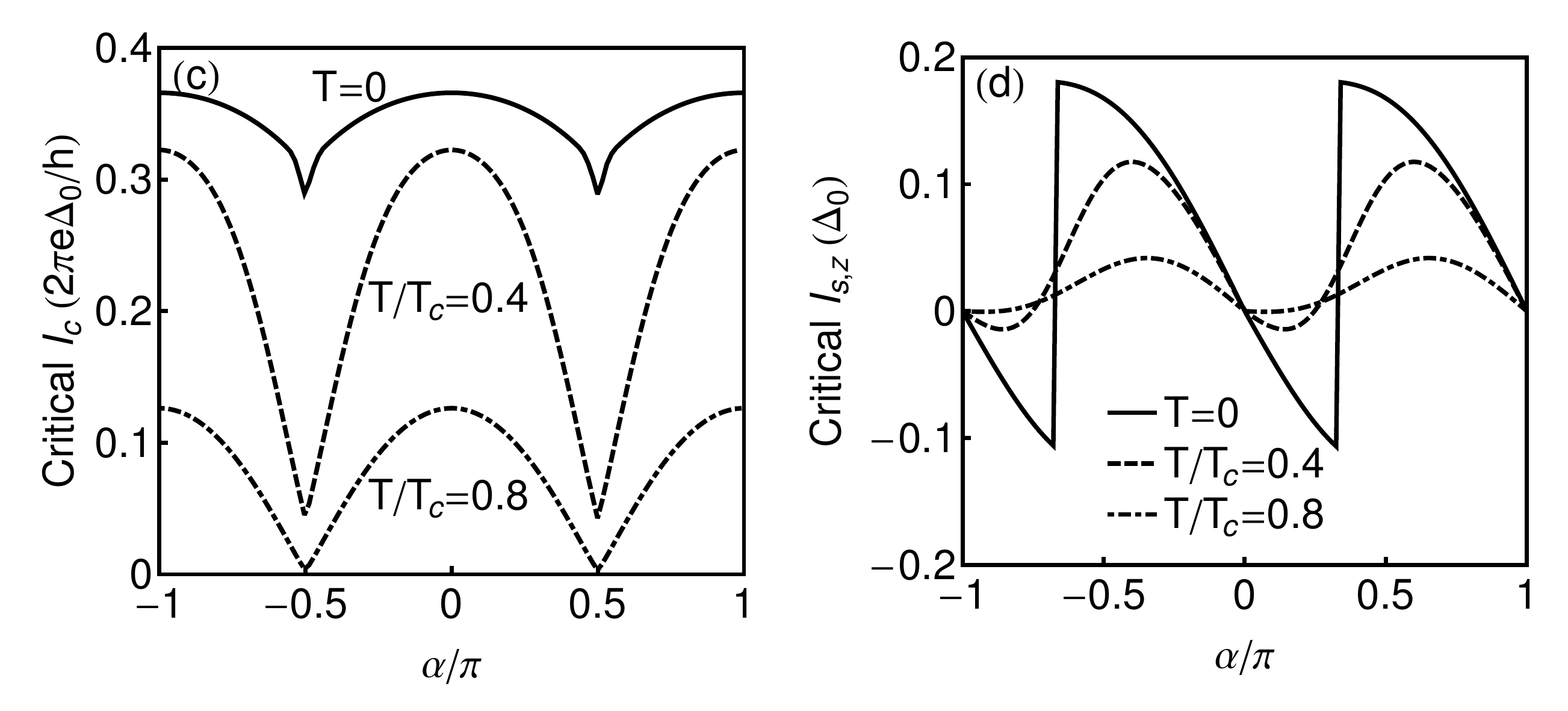}
\caption{(a) Critical Josephson charge and (b) spin current as a
  function of 
  temperature, at several different values of $\alpha$. (c) Critical Josephson
  charge and (d) spin current as a function of $\alpha$, at several
  different temperatures. In all panels we take $g=1$ and $Z=0$.} 
\label{fig:charge_spin_current_alpha_T}
\end{figure}

The critical charge current, defined as the maximum current with
respect to $\phi$, is a readily-accessible experimental quantity. We plot this
alongside the critical spin current
in~\fig{fig:charge_spin_current_alpha_T} as 
a function of the  temperature and the angle $\alpha$. As can be
seen in panel (a), the temperature-dependence of the critical charge current
qualitatively 
changes with the orientation of the magnetization: For $\alpha=0$ it grows
linearly with decreasing temperature immediately below $T_c$, and saturates at
$T\approx 0.2T_c$; In contrast, when $\alpha=0.5\pi$, it grows superlinearly
with decreasing $T$, but remains much smaller than the $\alpha=0$ current
until $T\approx 0.2T_c$, below which it displays a rapid
increase. This ``low-temperature anomaly'' in the $\alpha=0.5\pi$ critical
charge current reflects the importance of resonant tunneling
through the zero-energy Andreev bound states far below
$T_c$.~\cite{lowtempanomaly}  
Similarly, in panel (c) we observe that near $T_c$ the critical current as a
function of $\alpha$ follows the perturbation predictions of
$\max\{|I_c|\}\propto |\cos(\alpha)|$; as the temperature is decreased towards
$T=0$, however, the critical current at $\alpha=\pm0.5\pi$ increases
due to the resonant tunneling, and there is hence overall relatively weak
dependence of the critical current on the magnetization orientation. 
The critical spin current [\fig{fig:charge_spin_current_alpha_T}(b), (d)] also 
shows strong temperature- and $\alpha$-dependence.

\section{Conclusions}~\label{sec:conclusions}

In this paper we have studied the unconventional Josephson charge and spin
currents in a TFS junction. Using complementary theoretical methods, we have
established that single-Cooper-pair tunneling currents are possible
when the magnetization 
of the FM has a component parallel to the ${\bf d}$ vector of the
TS, and when the orbital pairing states of the superconductors
have the same parity with respect to the interface momentum. 
We hence see that spin and orbital degrees of freedom both play a critical
role in this junction; this is also the case when a lowest-order Josephson
effect between a TS and an SS is mediated by spin-orbit coupling at the
barrier.~\cite{AsaTanSigKas,GeshLark1986,Fenton,MilRaiSau1988} At a
microscopic level, the perturbation theory analysis reveals that the 
spin-dependent phase shifts of the tunneling Cooper pairs are responsible for
the charge and spin Josephson currents, due to the
interference of the spin-$\uparrow$ and spin-$\downarrow$ particle currents in
the TS. Surprisingly, the interference of the particle currents is also
responsible for a phase-dependent spin current in the TS, even though spin
currents are forbidden in the SS. Similar interference 
effects occur in the TFT junction,~\cite{BryMan2009} and junctions between
TSs with misaligned ${\bf d}$-vectors.~\cite{LinSudFMTS,asanospin}

Our analysis has confirmed previous observations of a highly unusual
charge current in the TFS junction.~\cite{TanKas1999,TanKas2000,YokTanGol2007}
Not only is it 
linearly proportional to 
the magnetization of the FM 
as $\propto {\bf M}\cdot{\bf d}$, but it also has unconventional $\cos\phi$
dependence on the phase difference (if time-reversal-symmetry is not
broken in one of the superconductors). This implies a contribution to the
junction free energy 
\beq
\propto {\bf M}\cdot{\bf d}\Delta_s\Delta_t\sin\phi \,. \label{eq:freeE}
\eeq
In this expression we can regard ${\bf d}\Delta_s\Delta_t\sin\phi$ as an
intrinsic interface spin which appears at junctions between a 
TS and an SS, even when the barrier is nonmagnetic. If the barrier is
ferromagnetic, the coupling of its magnetic moment to this intrinsic spin
therefore generates the lowest-order Josephson effect. Remarkably, such an
intrinsic interface spin is indeed known to exist in nonmagnetic TS-SS
junctions.~\cite{KwoSenYak2004,LuYip2009,SenYak2008,ZhaChaLinWan2011} For an
$s$-wave SS, this spin only appears for exactly the same $p$-wave TS orbital
configurations which would allow a lowest-order Josephson current in the
corresponding TFS junction.

The unusual form of the Josephson current in the TFS offers strong tests
for a triplet state. For instance, the observation of the linear dependence
of the current on ${\bf M}$ would be clear evidence of triplet pairing. On
the other hand, a domain structure in the ferromagnet could significantly 
reduce the Josephson current, as the currents across domains with opposite
magnetization would have opposite sign. This can be turned to our advantage,
however, as a magnetic flux line trapped at the boundary between two such
domains would be quantized in half-integer multiples of
$\Phi_{0}$. This
is a key signature of the lowest-order Josephson coupling in the TFS junction,
and could be directly imaged with SQUID microscopy, or deduced
from the Fraunhofer pattern.
More speculatively, the coupling~\eq{eq:freeE} could
spontaneously induce a magnetization in a barrier sufficiently close to a
magnetic instability, if the free energy gain due to the Josephson
coupling can offset the cost of magnetic
energy.~\cite{BryIniManSig2010,Arahata2013}

In our study we have neglected the likely variation of the superconducting
order parameter close to the junction interface. Since our results depend only
on the bulk properties of the superconductors, however, we do not expect
qualitative modification of our results. A more serious limitation of our
calculation is that we have not accounted for the torque exerted by the spin
current on the barrier's magnetic moment. Regarding the ${\bf d}$-vector as
fixed, we anticipate that the polarization of the spin current $\propto {\bf
  d}\times{\bf M}$ would cause a precession of the magnetization about the
${\bf d}$-vector, with eventual decay into the stable
configuration.~\cite{TSFM} If the ${\bf d}$-vector is only
weakly pinned, on the other hand, there may be a significant reconstruction of
the TS pairing 
state close to the interface. Although this is a very interesting problem, it
is beyond the scope of the current paper.

\acknowledgments

The authors acknowledge useful discussions with G. Annunziata,
M. Sigrist, Y. Tanaka, C. Timm, V. M. Yakovenko, and I. \v{Z}uti\'c.
Y.A. was supported by KAKENHI on Innovative Areas 
``Topological Quantum Phenomena'' (No. 22103002) from MEXT of Japan.


\begin{thebibliography}{99}

\bibitem{SigUed1991}M. Sigrist and K. Ueda, Rev. Mod. Phys. {\bf 63}, 239
  (1991). 

\bibitem{JoyTai2002}R. Joynt and L. Taillefer, Rev. Mod. Phys. {\bf 74},
  235 (2002).

\bibitem{Norman2011}M. R. Norman, Science {\bf 332}, 196 (2011).

\bibitem{MacMae2003}A. P. Mackenzie and Y. Maeno, Rev. Mod. Phys. {\bf 75},
  657 (2003).

\bibitem{Maeno2012}Y. Maeno, S. Kittaka, T. Nomura, S. Yonezawa,and
  K. Ishida, J. Phys. Soc. Jpn. {\bf 81}, 011009 (2012).

\bibitem{KalBer2009}C. Kallin and A. J. Berlinsky, J. Phys.: Condensed Matter
  {\bf 21}, 164210 (2009).


\bibitem{tunspec}M. Yamashiro, Y. Tanaka, and S. Kashiwaya, Phys. Rev. B
  {\bf 56}, 7847 (1997); K. Sengupta, H.-J. Kwon, and V. M. Yakovenko,
  Phys. Rev. B {\bf 65}, 104504 (2002); Y. Tanuma, Y. Tanaka, and
  S. Kashiwaya, Phys. Rev. B {\bf 74}, 024506 (2006); S. Wu and K. V. Samokhin,
  Phys. Rev. B {\bf 81}, 214506 (2010).

\bibitem{TSDN}Y.\ Tanaka and S.\ Kashiwaya, Phys.\ Rev.~B \textbf{70}, 012507
  (2004); Y.\ Asano, Y.\ Tanaka, and S.\ Kashiwaya,
  Phys.\ Rev.\ Lett.\ \textbf{96}, 097007 (2006); Y. Asano, A. A. Golubov,
  Y. V. Fominov, and Y. Tanaka, Phys. Rev. Lett. \textbf{107}, 087001 (2011). 

\bibitem{tunspecTSFM}T. Hirai, Y. Tanaka, N. Yoshida, Y. Asano, J. Inoue, and
  S. Kashiwaya, Phys. Rev. B {\bf 67}, 174501 (2003).

\bibitem{Annunziata2011}G. Annunziata, M. Cuoco, C. Noce, A. Sudb\o, and
  J. Linder, Phys. Rev. B {\bf 83}, 060508(R) (2011); G. Annunziata,
    D. Manske, and J. Linder, Phys. Rev. B {\bf 86}, 174514 (2012).

\bibitem{TSFM}P. M. R. Brydon, Phys. Rev. B {\bf 80}, 224520 (2009);
  P. M. R. Brydon, Y. Asano, and C. Timm, Phys. Rev. B {\bf 83}, 180504(R)
  (2011). 

\bibitem{TSFMprox}P. Gentile, M. Cuoco, A. Romano, C. Noce, D. Manske, and
  P. M. R. Brydon, arXiv:1208.5871; D. Terrade, P. Gentile, M. Cuoco, and
  D. Manske, arXiv:1210.5160. 

\bibitem{FTFspinvalve}F. Romeo and R. Citro, arXiv:1303.0375.

\bibitem{asanospin} Y.~Asano, Phys. Rev. B \textbf{72}, 092508 (2005);
  \textbf{74}, 220501(R) (2006). 

\bibitem{LinSudFMTS}M. S. Gr{\o}nsleth, J. Linder, J.-M. B{\o}rven, and
  A. Sudb\o, Phys. Rev. Lett. {\bf 97}, 147002 (2006); J. Linder,
  M. S. Gr{\o}nsleth, and A. Sudb\o, Phys. Rev. B {\bf 75}, 024508 (2007).

\bibitem{fomin}I.\ A. Fomin, Zh. Eksp. Teor. Fiz. {\bf 88}, 2039 (1985) [Sov.
Phys. JETP {\bf 61}, 1207 (1985)].

\bibitem{bunkov} Y.\ M.\ Bunkov, V.\ V.\ Dmitriev, A.\ V.\ Markelov, 
  Y.\ M.\ Mukharskii, and D. Einzel, Phys. Rev. Lett. \textbf{65}, 867
  (1990). 

\bibitem{Kastening06}B. Kastening, D. K. Morr, D. Manske, and K. Bennemann,
  Phys. Rev. Lett. {\bf 96}, 047009 (2006).

\bibitem{Brydon08} P. M. R. Brydon, B. Kastening, D. K. Morr, and D. Manske,
  Phys. Rev. B {\bf 77}, 104504 (2008)
  
\bibitem{BryMan2009}P. M. R. Brydon and D. Manske, Phys. Rev. Lett. {\bf 103},
  147001 (2009).

\bibitem{BujTimBry2012}B. Bujnowski, C. Timm, and P. M. R. Brydon, J. Phys.:
  Condensed Matter {\bf 24}, 045701 (2012).

\bibitem{GolKupIli2002}A. A. Golubov, M. Y. Kupriyanov, and E. Il'ichev,
  Rev. Mod. Phys. {\bf 76}, 411 (2002).

\bibitem{Eschrig2003}M. Eschrig, J. Kopu, J. C. Cuevas, and G. Sch\"on,
  Phys. Rev. Lett. {\bf 90}, 137003 (2003).

\bibitem{SSFMreviews}A. I. Buzdin, Rev. Mod. Phys. {\bf 77}, 935 (2005);
  F. S. Bergeret, A. F. Volkov, and K. B. Efetov, Rev. Mod. Phys. {\bf 77},
  1321 (2005). 


\bibitem{krockenberger}  Y.\ Krockenberger, M.\ Uchida, K.\ S.\ Takahashi, 
M.\ Nakamura, M.\ Kawasaki, and Y.\ Tokura, Appl. Phys. Lett. \textbf{97}, 082502 (2010).


\bibitem{PalHaeMaa1977}J. A. Pals, W. van Haeringen, and M. H. van Maaren,
  Phys. Rev. B {\bf 15}, 2592 (1977).

\bibitem{Yip1993} S. Yip, J. Low. Temp. Phys. {\bf 91}, 203 (1993).

\bibitem{KwoSenYak2004}H.-J. Kwon, K. Sengupta, and V. M. Yakovenko,
  Eur. Phys. J. B {\bf 37}, 349 (2004).

\bibitem{LuYip2009}C.-K. Lu and S. Yip, Phys. Rev. B {\bf 80}, 024504 (2009).

\bibitem{Fenton} E. W. Fenton, Solid State Commun. {\bf 34}, 917 (1980); {\bf
  54}, 709 (1985); {\bf 60}, 347 (1986).

\bibitem{GeshLark1986} V. B. Geshkenbein and A. I. Larkin, Pis'ma
  Zh. Eksp. Teor. Fiz. {\bf 43}, 305 (1986) [JETP Lett. {\bf 43},
  395 (1986)].

\bibitem{MilRaiSau1988}A. Millis, D. Rainer, and J. A. Sauls, Phys. Rev. B
  {\bf 38}, 4504 (1988).

\bibitem{AsaTanSigKas}Y. Asano, Y. Tanaka, M. Sigrist, and S. Kashiwaya,
  Phys. Rev. B {\bf 67}, 184505 (2003); {\bf 71}, 214501 (2005). 

\bibitem{ZutMaz2005}I. \v{Z}uti\'{c} and I. Mazin, Phys. Rev. Lett. {\bf 95},
  217004 (2005).

\bibitem{Jin2000}R. Jin, Y. Liu, Z. Q. Mao, and Y. Maeno, Europhys. Lett. {\bf
  51}, 341 (2000).


\bibitem{Nelson2004}K. D. Nelson, Z. Q. Mao, Y. Maeno, and Y. Liu, Science
  {\bf 306}, 1151 (2004).


\bibitem{Liu2010}Y. Liu, New. J. Phys. {\bf 12}, 075001 (2010).

\bibitem{Saitoh2012}K. Saitoh, S. Kashiwaya, H. Kashiwaya, M. Koyanagi,
  Y. Mawatari, Y. Tanaka, and Y. Maeno, Appl. Phys. Express {\bf 5}, 113101
  (2012). 

\bibitem{TanKas1999}Y. Tanaka and S. Kashiwaya. J. Phys. Soc. Jpn. {\bf 68},
  3485 (1999).

\bibitem{TanKas2000}Y. Tanaka and S. Kashiwaya, J. Phys. Soc. Jpn. {\bf 69},
  1152 (2000).  

\bibitem{YokTanGol2007}T. Yokoyama, Y. Tanaka, and A. A. Golubov, Phys. Rev. B
  {\bf 75}, 094514 (2007).

\bibitem{BruOttZim1995}C. Bruder, A. van Otterlo, and G. T. Zimanyi,
  Phys. Rev. B {\bf 51}, 12904(R) (1995).

\bibitem{Mahan}G. D. Mahan, {\it Many-Particle Physics} (Kluwer Academic, New
  York, 2000).

\bibitem{Lee}P.~A.\ Lee and D.~S.\ Fisher, Phys. Rev. Lett. \textbf{47}, 882
  (1981). 

\bibitem{asano01}Y.~Asano, Phys. Rev. B \textbf{63}, 052512 (2001).

\bibitem{FurTsu1991}A. Furusaki and M. Tsukada, Solid State Commun. {\bf
    78}, 299 (1991). 


\bibitem{KasTan2000}S. Kashiwaya and Y. Tanaka, Rep. Prog. Phys. {\bf 63},
  1641 (2000).

\bibitem{currentformula}C. W. J. Beenakker and H. van Houten, in {\it
  Nanostructures and Mesoscopic Systems}, edited by W. P. Kirk and M. A. Reed
  (Academic, New York, 1992).

\bibitem{lowtempanomaly} Y.~Tanaka and S.~Kashiwaya, Phys. Rev. B \textbf{53},
  11957(R) (1996); Y.~S.~Barash, H.~Burkhardt, and D.~Rainer,
  Phys. Rev. Lett. \textbf{77}, 4070 (1996). 

\bibitem{SenYak2008}K. Sengupta and V. M. Yakovenko, Phys. Rev. Lett. {\bf
  101}, 187003 (2008).

\bibitem{ZhaChaLinWan2011}H. Zhang, K. S. Chan, Z. Lin, and J. Wang, J. Phys:
  Condensed Matter {\bf 23}, 415701 (2011).

\bibitem{BryIniManSig2010}P. M. R. Brydon, C. Iniotakis, D. Manske, and
  M. Sigrist, Phys. Rev. Lett. {\bf 104}, 197001 (2010).

\bibitem{Arahata2013}E. Arahata, T. Neupert, and M. Sigrist, Phys. Rev. B {\bf
  87}, 220504(R) (2013).


\end{thebibliography}
\end{document}